\documentclass[a4paper,11pt]{article}
\pdfoutput=1 

\usepackage{jcappub} 

\usepackage[T1]{fontenc} 

\title{\boldmath On Magnetized  Neutron Stars}

\author[a,b,1]{Luiz Lopes,\note{Corresponding author.}}
\author[b]{Debora Menezes}

\affiliation[a]{ Centro Federal de Educa\c{c}\~ao Tecnol\'ogica de Minas Gerais Campus VIII; CEP 37.022-56, Varginha - MG - Brasil
}
\affiliation[b]{ Departamento de Fisica, CFM - Universidade Federal de Santa Catarina;  C.P. 476, CEP 88.040-900, Florian\'opolis, SC, Brasil 
}

\emailAdd{llopes@varginha.cefetmg.br}
\emailAdd{debora.p.m@ufsc.br}

\abstract{In this work we review the  formalism normally used in the literature
about the effects of density-dependent magnetic fields
 on the properties of neutron  and quark stars, expose some ambiguities that
 arise and propose a way to solve the related problem.
  Our approach explores more deeply the concept of 
  pressure, yielding the so called chaotic magnetic field formalism for the stress
  tensor. We  also  use a different way of introducing a variable magnetic
  field, which depends on the energy density rather than on the
  baryonic density, which allows us to build a parameter free model.}

\arxivnumber{1411.7209}

\begin{document}
\maketitle
\flushbottom

\section{Introduction}
\label{sec:intro}

Neutron stars are  objects with densities much higher than those found
in terrestrial laboratories, what makes them a 
valorous subject of study.  It has been suggested that a possible
 source of anomalous X-ray pulsars and soft gamma-ray repeaters, is
 the decay of very high magnetic fields, which powers these objects. In this case,
the magnetic field in the surface of some neutron stars could be much larger than it was previously thought, reaching values  as strong as $10^{14} - 10^{15}$ G.
These objects are called magnetars~\cite{Duncan1,Duncan2,Duncan3,Usov,Hurley}. Although fields of the order of $10^{15}$ G do not affect
the main properties of neutron stars, fields larger than $10^{18}$G are expected in the neutron star core due to the scalar Virial
theorem~\cite{Teu}. To simulate the variation of the magnetic field with the density, an $\it{ad~hoc}$ exponential density-dependent
magnetic field was proposed in ref.~\cite{Pal} and widely adopted in subsequent works~\cite{Mao,Rabhi,Menezes1,Ryu,Rabhi2,Mallick,Lopes1,Dex,Benito1,Benito2,Mallick2,Ro,Dex2}. 

The main problem found is that the formalism commonly used  does not  seems to be  precise. As pointed out
by two classical books of gravitation~\cite{Misner,Zel}, when anisotropies are present, the concept of pressure must be treated with more care.
Moreover, macroscopic properties of the neutron stars, as mass and radii, now depend not only on the strength of the magnetic fields,
but also on the way in which it varies with the density. However,
since it was introduced in an $\it{ad~hoc}$ way, the 
macroscopic properties depend
on arbitrary free parameters, compromising the accuracy of the results. 

In this work we try to fix these issues introducing the chaotic magnetic field approximation~\cite{Zel}. Within this approach,
the contribution of the magnetic field  yields the well-known
radiation pressure,   circumventing  the problem of anisotropies.
Also, we propose that the magnetic field be coupled to the energy
density rather than to the number density.
This seems a more natural approach since it is the energy density,
instead of the number density
 that determines the macroscopic quantities mass and radius in the TOV equations~\cite{TOV}.
With this { new assumption}, the number of free parameters is reduced from two to only one ($\gamma$). Moreover, as we will see, 
 within the chaotic field approximation, for $\gamma$ $\ge$ 2.0, we
 obtain a parameter free model!
The effects of magnetic fields in neutron stars are discussed within
two possible star configurations: with and without 
hyperons.  The probable existence of hyperons in neutron star interior is an old
subject of study~\cite{Ambar} but still a very active field of research~\cite{Glen,Paoli,Weiss1,Weiss2,Lopes2}.

 We also discuss brielfly quark star properties within the chaotic magnetic field, to study its effects on models besides QHD.
Quark stars can reach values of magnetic fields higher than in neutron
stars composed of hadronic matter, because they are
bound by the strong force instead of the gravity and hence, they are a
good source for our investigation.

This works is organized as follows. We first discuss briefly the
formalism and parameter\-ization of nuclear matter subject to
strong magnetic fields,
and present the current proposal of density dependent magnetic field. Then we discuss the ambiguities due to the anisotropy and the  free parameters, 
and how much they  influence  the macroscopic properties of neutron stars. After, we introduce our proposal within the  chaotic field approximation, 
showing how this formalism avoids the problem of anisotropies, and then  present our model with the energy density dependent magnetic field. We prove that this proposal contributes  to a parameter free
model and helps with the puzzle of { small} neutron star radii. Finally we discuss quark stars with magnetic fields.
 At the end, the conclusions of  the present work are drawn.

\section{Current Hadronic Formalism for Neutron Stars}

We use an extended version of the relativistic QHD~\cite{Serot}, whose Lagrangian density reads:
\begin{eqnarray}
\mathcal{L}_{QHD} = \sum_b \bar{\psi}_b \bigg [\gamma^\mu(i\partial_\mu -e_bA_\mu - g_{b,v}\omega_\mu  - g_{b,\rho} \frac{1}{2}\vec{\tau} \cdot \vec{\rho}_\mu)
- (m_b - g_{b,s}\sigma ) \bigg ]\psi_b     + \frac{1}{2} m_v^2 \omega_\mu \omega^\mu 
   \nonumber \\ + \frac{1}{2} m_\rho^2 \vec{\rho}_\mu \cdot \vec{\rho}^{ \; \mu}   + \frac{1}{2}(\partial_\mu \sigma \partial^\mu \sigma - m_s^2\sigma^2)  
    - U(\sigma)  - \frac{1}{16\pi}F^{\mu \nu}F_{\mu \nu}  - \frac{1}{4}\Omega^{\mu \nu}\Omega_{\mu \nu} -  \frac{1}{4}\bf{P}^{\mu \nu} \cdot \bf{P}_{\mu \nu}  , \label{s1} 
\end{eqnarray}
in natural units. The sum in $b$ stands just for the nucleons or for all the baryon octet, depending on our choice for the star constituents,
 $\psi_b$  are the    Dirac fields of the baryons,  $\sigma$, $\omega_\mu$ and $\vec{\rho}_\mu$ are the mesonic fields,
 and $A_\mu$ is the electromagnetic four-potential.
 The $g's$ are the Yukawa coupling constants that simulate the strong interaction, $m_b$ and $e_b$  are the mass and the electric
charge  of the baryon $b$; $m_s$, $m_v$, and $m_\rho$ are
 the masses of the $\sigma$, $\omega$,  and $\rho$ mesons respectively.
 The antisymmetric  field  tensors are given by their usual expressions as presented in~\cite{Glen}.
  The $U(\sigma)$ is the self-interaction term introduced in ref.~\cite{Boguta} to fix some of the saturation properties of the nuclear matter.
We also define $M^{*}_b$ as the effective mass of the baryon $b$: $M^{*}_b = M_b - g_{sb}\sigma$.

In the presence of a magnetic field $B$ in the $z$ direction, the energy eigenvalue $E_b$,
and the number density $n_b$ of charged baryons are quantized:

\begin{equation}
E_b = \sqrt{M^{*2}_b - k_z^2 + 2\nu|e|B} , \quad n_b = \sum_\nu \frac{|e|B}{2\pi^2} k_z \label{s2},
\end{equation}
where the discrete parameter $\nu$ is called Landau level $(LL)$. The uncharged baryons keep their 
usual expressions~\cite{Glen}. The mesonic fields are obtained by mean
field approximation~\cite{Serot,Glen,Lopes1} 
and the equation of state (EoS)  by thermodynamic
relations~\cite{Greiner}. To construct a 
$\beta$ stable matter, we also 
include leptons as free Fermi gas and impose zero net charge and chemical equilibrium.

To describe the properties of  nuclear matter we use the well-known GM1 parametrization~\cite{Glen2},
a widely accepted parametrization~\cite{Lopes1,Benito1,Benito2,Glen,Paoli,Weiss1,Weiss2,Lopes2} that is able to reasonably describe both, nuclear matter
and stellar structure, consistent with experimental and astrophysical observations~\cite{Lopes2}. Another parametrization would predict different 
values for the macroscopic properties of the neutron stars, as masses and radii, nevertheless we do not expect that it would change our conclusion
about the effects of the magnetic field in neutron star properties.

To fix the hyperon-meson coupling constant, we follow ref.~\cite{Lopes2}, which use a complete SU(3) model to fix
 all meson-baryon interaction. Moreover, the vector mesons are fixed within a more restrictive SU(6) parametrization,
while the scalar mesons are fixed within a {\it nearly} SU(6) parametrization. In other words, the hyperon-meson
coupling constants are:

\begin{eqnarray}
\frac{g_{\Lambda\omega}}{g_{N\omega}} = \frac{g_{\Sigma\omega}}{g_{N\omega}} = 0.667,  \quad \frac{g_{\Xi\omega}}{g_{N\omega}} =  0.333, \nonumber \\
\frac{g_{\Sigma\rho}}{g_{N\rho}} = 2.0 \quad \frac{g_{\Xi\rho}}{g_{N\rho}} = 1.0 , \quad \frac{g_{\Lambda\rho}}{g_{N\rho}} = 0.0,  \\ \label{s3}
\frac{g_{\Lambda\sigma}}{g_{N\sigma}} = 0.610 , \quad \frac{g_{\Sigma\sigma}}{g_{N\sigma}} =  0.396 , \quad \frac{g_{\Xi\sigma}}{g_{N\sigma}} = 0.113 .\nonumber
\end{eqnarray}

\subsection{Standard density-dependent magnetic field}

In the current literature, the contribution of the electromagnetic field $(B^2/8\pi)$ is directly summed to the EoS to give 
the total energy density and
pressure~\cite{Rabhi,Menezes1,Ryu,Lopes1,Dex,Benito1,Benito2,Ro,Dex2, Prakash} as:

\begin{equation}
 \epsilon_T =  \epsilon_M + \frac{B^2}{8\pi}; \quad P_T = P_M + \frac{B^2}{8\pi} , \label{s4}
\end{equation}
where the subscript $M$ stands for the matter contribution for the EoS. 
To simulate the variation of the magnetic field with the density, an $\it{ad~hoc}$ exponential density-dependent
magnetic field is { normally} utilized~\cite{Pal,Mao,Rabhi,Menezes1,Ryu,Rabhi2,Mallick,Lopes1,Dex,Benito1,Benito2,Mallick2,Ro,Dex2}:

\begin{equation}
 B(n) = B^{surf} + B_0\bigg [ 1 - \exp \bigg \{ - \beta \bigg ( \frac{n}{n_0} \bigg )^{\theta} \bigg \} \bigg ], \label{s5} 
\end{equation}
where  $B^{surf}$  is the magnetic field on the surface of the neutron stars, taken as $10^{14}G$, $n$ is the total number
density, and $n_0$ is the nuclear saturation density.  Then $B$ is replaced by $B(n)$ in the term $B^2 /8\pi$ in
the EoS.

From Eq. (\ref{s4}) a large ambiguity arises immediately. Since neither the Lagrangian (Eq. (\ref{s1})) nor the astronomical observation contain
 information about how the magnetic field vary in neutron star interiors, any values  of the non-observables parameters
 $\beta$ and $\theta$ are equally valid. Indeed, in the literature there are several  sets for theses parameters. We analyse how some of 
them affect the macroscopic properties of neutron stars. The chosen ones are presented in Table~\ref{T1}.

\begin{table}[ht]
\begin{center}
\begin{tabular}{|c|cc|c|}
\hline 
   Set  &  $\beta$ & $\theta$ & From  \\
 \hline
 I   &  -  & -  &  $B = 0 G$  \\
 
  II & $1.0 \times 10^{-2}$ & 3.0 & Ref.~\cite{Pal} \\
 
  III & $5.0 \times 10^{-2}$ & 2.0 & Ref.~\cite{Rabhi} \\
 
  IV  & $5.0 \times 10^{-5}$ & 3.0 & Ref.~\cite{Menezes1} \\

 V & $1.0 \times 10^{-1}$ & 1.0 & Ref.~\cite{Mallick} \\

 VI & $2.0 \times 10^{-1}$ & 2.0 & Ref.~\cite{Mallick} \\
  
 VII & $6.5 \times 10^{-3}$ & 3.5 & Ref.~\cite{Lopes1} \\
\hline
\end{tabular} 
\caption{Different values for the non-observables parameters $\beta$ and $\theta$. We also
compare with the zero magnetic field approximation.} 
\label{T1}
\end{center}
\end{table}

\begin{figure}[tbp]
\begin{tabular}{cc}
\centering 
\includegraphics[width=.33\textwidth,origin=c,angle=270]{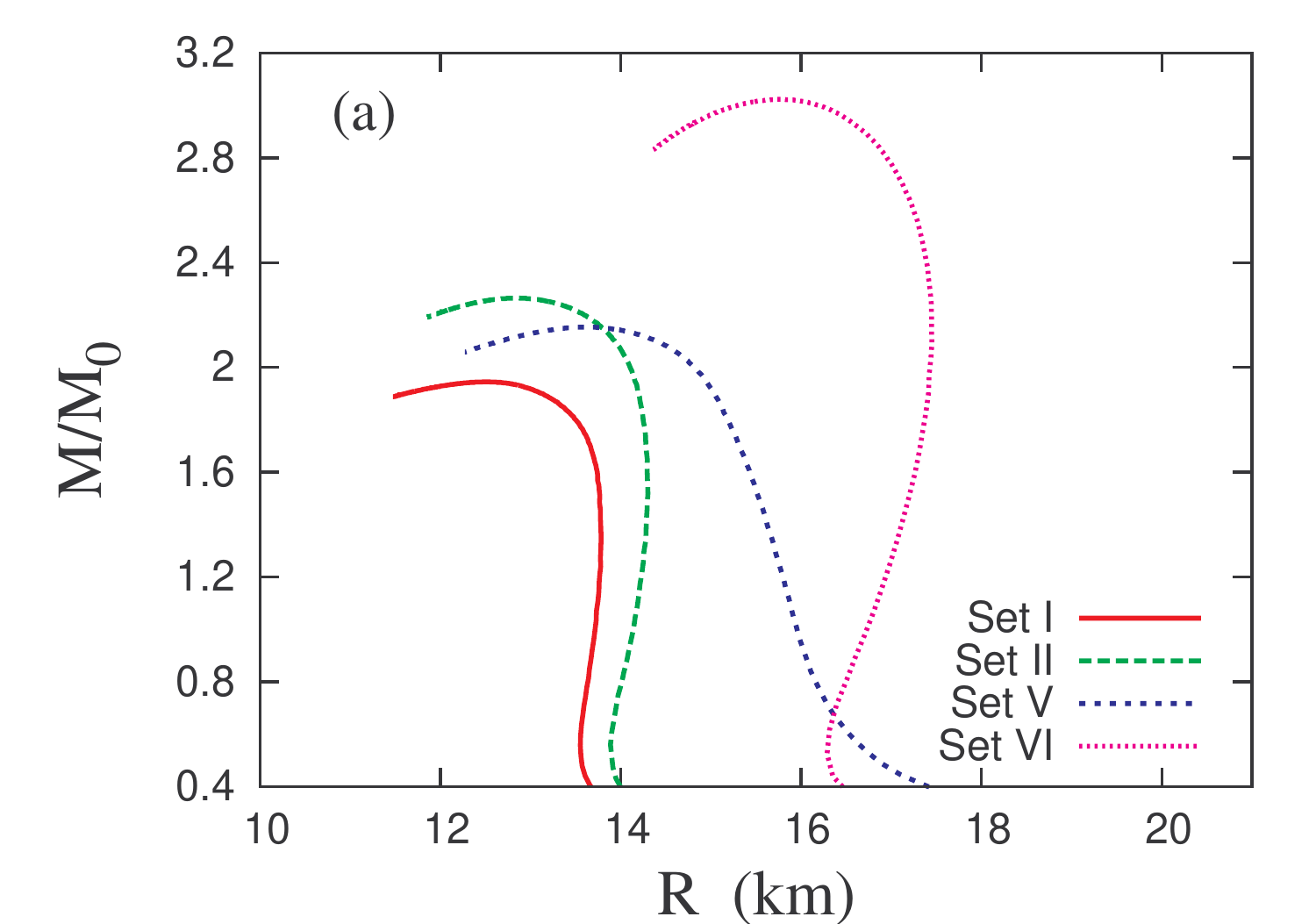} &
\includegraphics[width=.33\textwidth,origin=c,angle=270]{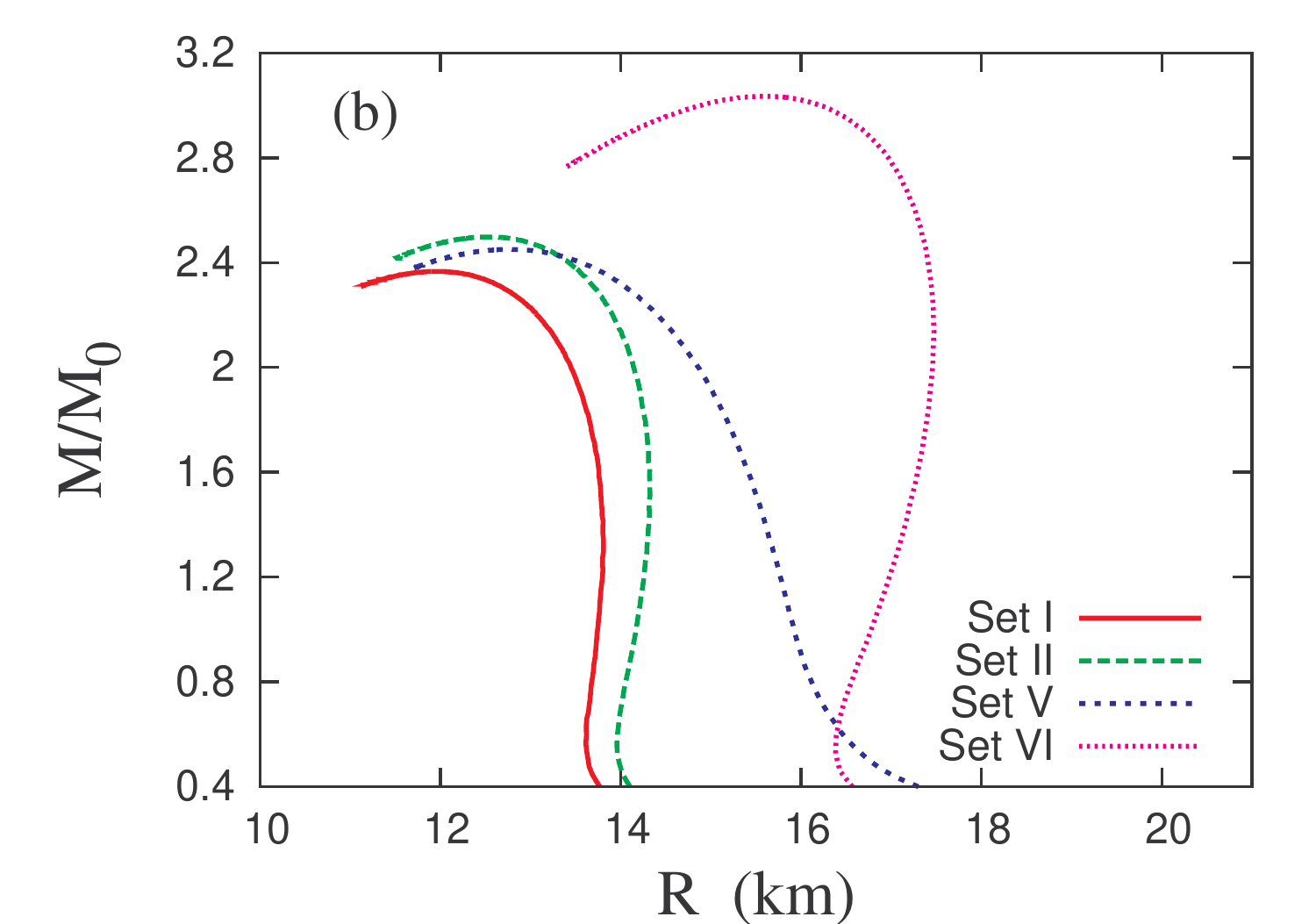}\\
\includegraphics[width=.33\textwidth,origin=c,angle=270]{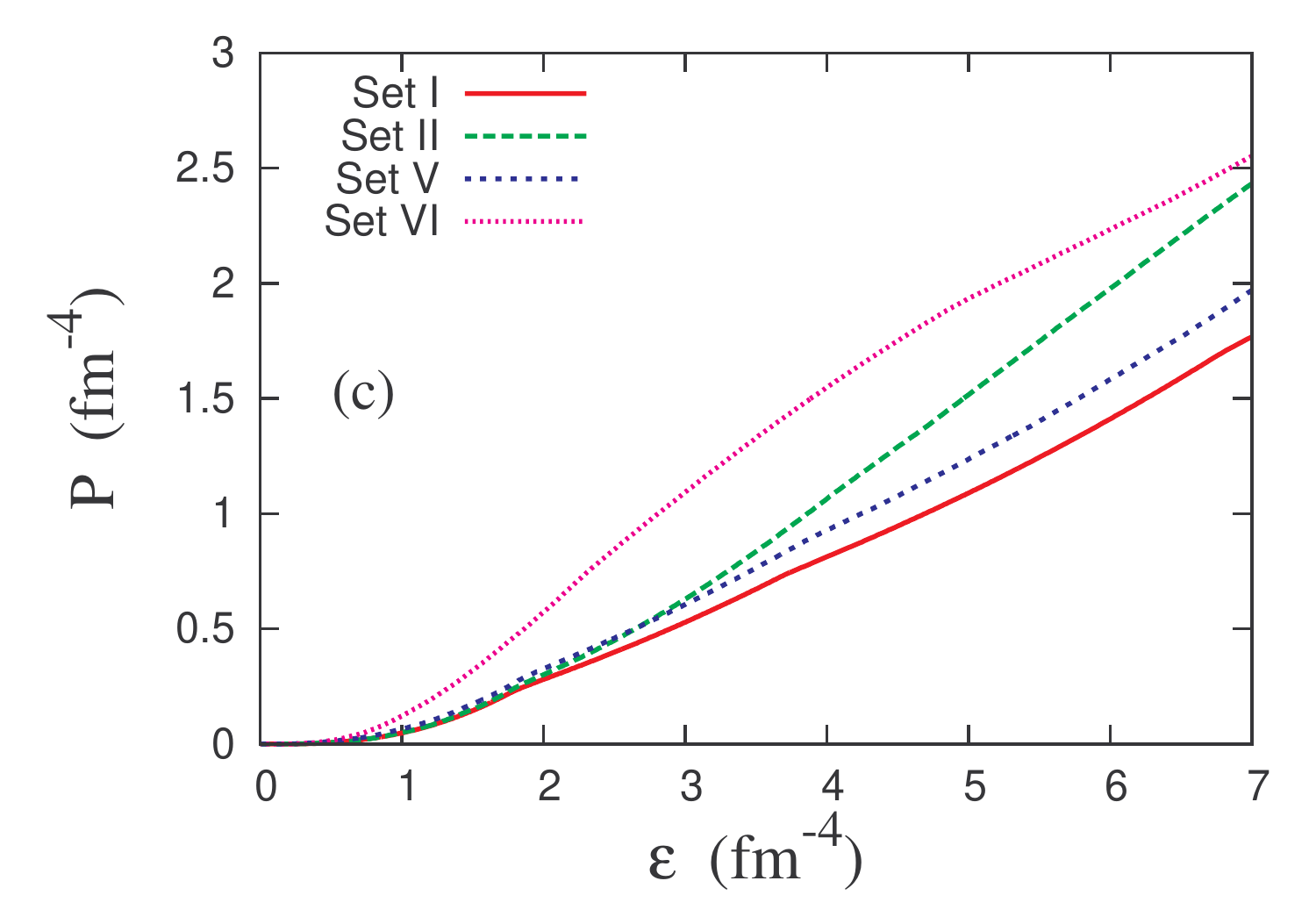} &
\includegraphics[width=.33\textwidth,origin=c,angle=270]{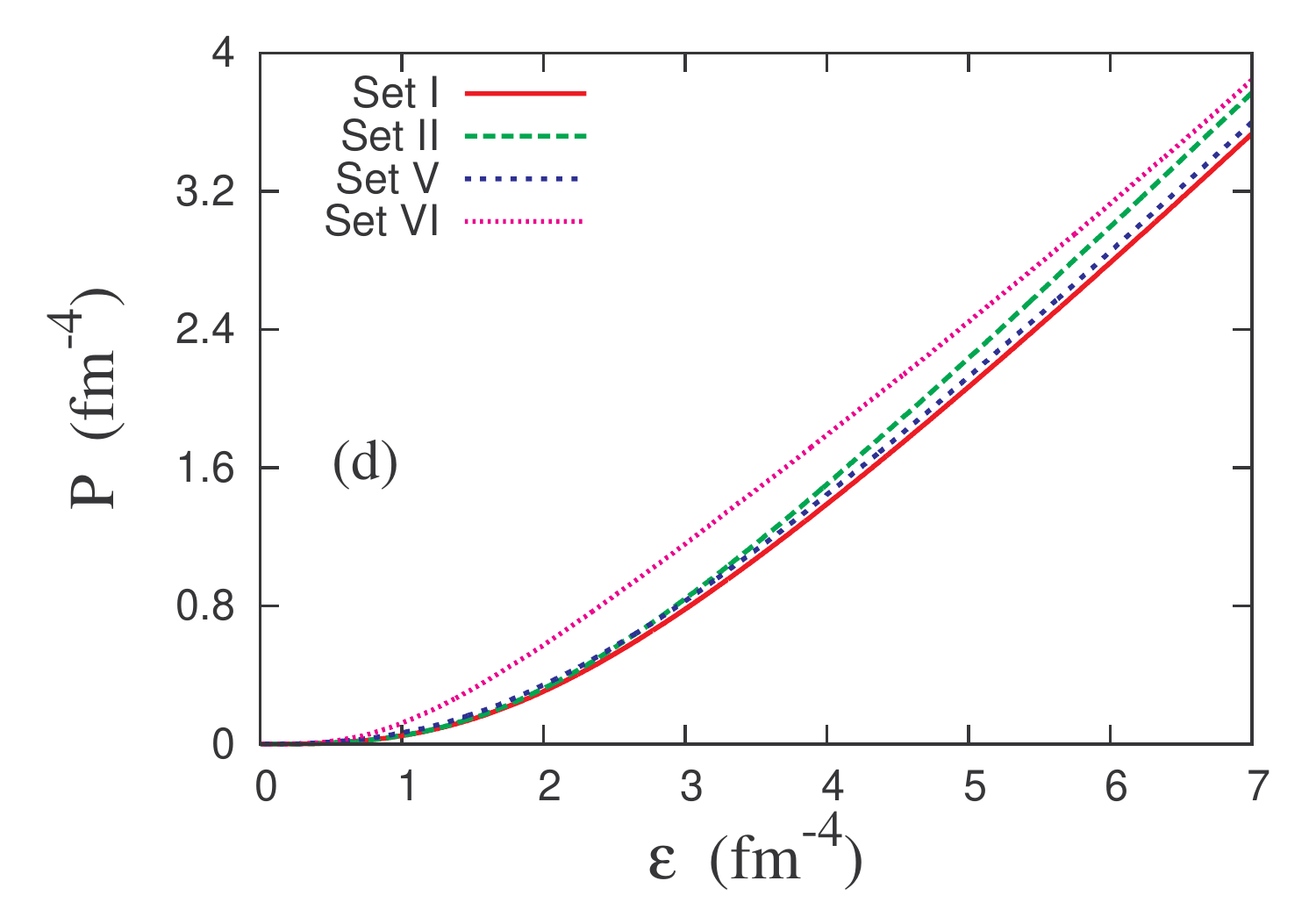}\\
\end{tabular}
\caption{\label{F1} (Color online)(Above) Mass-radius relation for neutron stars with (a) and without (b) hyperons,
and (below) their correspondent EoS (c) and (d). The influence of the
non-observables parameters $\beta$ and 
$\theta$ can increase
 the mass { by} more than $50\%$.}
\end{figure}

The validity of the results depends on the strength of the magnetic field.
Ref.~\cite{Xu1} shows that the EoS can be treated as practically isotropic  for fields up to 3.1 $\times 10^{18}G$, and
a more recent work~\cite{Xu2} corroborates this result. As in ref.~\cite{Lopes1,Benito1,Benito2}, we follow this prescription and  utilize the 
value of 3.1 $\times 10^{18}G$ as an upper limit for the magnetic field,
 although we can find fields as high as $10^{19}G$ in recent works~\cite{Dex,Ro}.
Now we solve the TOV equations~\cite{TOV} using   the EoS from Eq.~(\ref{s4}) as input. 
The mass-radius relation for neutron stars with and without hyperons and
their corresponding EoS 
are plotted in  figure~\ref{F1} for some sets of $\beta$ and $\theta$.

\begin{table}[ht]
\begin{center}
\begin{tabular}{|c|ccc|ccc|}
\hline
& \multicolumn{3}{c|}{ Hyperons included} & \multicolumn{3}{c|}{Nucleons only} \\
\hline 
   Set  &  $M/M_\odot$ & R  (km)  & $\epsilon_c$ ($fm^{-4}$) & $M/M_\odot$ & R  (km)  & $\epsilon_c$ ($fm^{-4}$)  \\
 \hline
 I   &  1.94  & 12.5  &  5.01 & 2.37 & 12.1 & 5.69  \\
 
  II & 2.15 & 12.1 & 5.52 & 2.44 & 12.1 & 5.48 \\
 
  III & 2.26 & 12.8 & 4.81  & 2.50 & 12.5 & 5.09 \\
 
  IV  & 1.94 & 12.5 & 4.97 & 2.37 & 12.0 & 5.63 \\

 V & 2.15 & 13.6 & 4.39 & 2.45 & 12.8 & 5.14 \\

 VI & 3.02 & 15.7 & 3.18 & 3.04 & 15.6 & 3.26 \\ 
  
 VII & 2.20 & 12.0 & 5.60 & 2.46 & 12.1 & 5.46 \\
\hline
\end{tabular} 
\caption{Maximum mass, the correspondent radius and the central energy density
 for different sets of $\beta$ and $\theta$ for neutron stars with and without hyperons.} 
\label{T2}
\end{center}
\end{table}

We also resume the properties of the maximum mass neutron star  in Table~\ref{T2} for all chosen sets.
We see that the magnetic field always hardens the EoS, increasing
the maximum possible mass. However, the inherent ambiguity present in the
non-observable parameters is strongly reflected  on the
maximum possible mass and the radius of the neutron stars.

Note that all results are obtained 
using the GM1 parametrization and   the same  hyperon-meson
coupling constant values.  Moreover,  the contribution of the magnetic field
to the matter,  the Landau quantization in the energy eigenvalue
and in the number density as expressed in the Eq.~(\ref{s2}), the number
of Landau levels and  $\epsilon_M$ and $P_M$ in Eq.~(\ref{s4}),
are calculated with a fixed value of the magnetic field; in our case
$B=3.1 \times 10^{18}G$; and having therefore, exactly the same values.
The variable magnetic field only affects the proper energy density and pressure,
 given by  the term $B^2/8\pi$.
 If someone wonders about how much the magnetic
field influences the EoS and the macroscopic properties of the neutron stars,
the most sincere answer  is we don't know, since we don't know
how the magnetic field varies inside the neutron star interior.
The  mass could vary from zero if we choose  
Set IV to $1.08M_\odot$ within the Set VI for 
a star with hyperons in the core. Even when the maximum mass is similar, as in  Sets
II and V, the value of the radius is significantly different. 
Also, within  Set VI, the EoS with and without hyperons
are almost degenerate, predicting maximum masses and radii
very close to each other,  and having their maximum masses very close
to the theoretical limit of 3.2$M_\odot$~\cite{Ruffini}. 
We can also vary the magnetic field in the matter ($\epsilon_M$ and $P_M$)
as in ref.~\cite{Benito1,Ro}, however we chose not to follow this way
since we want to emphasize the influence of the term $B^2/8\pi$ itself.
Moreover, ref.~\cite{Benito1} shows that the results do not change
considerably by maintaining or not the density dependence of the magnetic field
applied to the matter.

In the appendix we make a deeper discussion about
the anisotropy and shear stress of the Maxwell's stress tensor.

\section{Chaotic Magnetic Field Formalism}

Now, we try to fix the severe ambiguity raised in the last section.
Our first query is about the formalism of directly adding the term
$B^2/8\pi$ in the pressure~\cite{Rabhi,Menezes1,Ryu,Lopes1,Dex,Benito1,Benito2,Ro,Dex2,Prakash},
as presented in Eq. (\ref{s4}).
This would be correct only if all components of the stress
tensor were equal, as pointed in ref.~\cite{Misner}, ``if they are not
identical, a rotation in the  frame of reference
will reveal the presence of shear stress (pag 140)'',
however, for a magnetic field in the $z$ direction,it is well-known that the
stress tensor has the form: diag$(B^2/8\pi,B^2/8\pi,-B^2/8\pi)$,
being non identical.
Ref.~\cite{Zel} go beyond and stands that in the presence of 
magnetic field the concept of pressure is lost. Nevertheless, they
give us a way to treat the effects of magnetic field:
``It is possible to describe the effect of the magnetic field by using the
pressure concept only when we are dealing with a small-scale
chaotic field (pag 158)''. In this case the stress tensor reads:
diag$(B^2/24\pi,B^2/24\pi,B^2/24\pi)$, avoiding the anisotropy problem, 
and yielding $P = \epsilon/3$, a radiation pressure formalism.
This also agrees with the field theory, where the pressure is
calculated as~\cite{Serot,Dex}: 

\begin{equation}
 P =  \frac{1}{3} < T_i^i > = \frac{1}{3} \bigg ( \frac{B^2}{8\pi} + \frac{B^2}{8\pi}  - \frac{B^2}{8\pi}  \bigg ) =  \frac{B^2}{24\pi} . \label{s6}
\end{equation}
even without the chaotic field approximation. However, we need to keep in mind that, despite the coincidence of values, Eq.~(\ref{s6}) does not 
represent the true thermodynamic pressure, since the components of the stress tensor are not equal~\cite{Zel}.

 Hence, to study the influence of the magnetic field on  neutron stars,
instead of using Eq.~(\ref{s4}), we have opted to use  Eq.~(\ref{s7}),
which seems more suitable:

\begin{equation}
\epsilon_T =  \epsilon_M + \frac{B^2}{8\pi}  , \quad P_T = P_M + \frac{B^2}{24\pi}.  \label{s7}
\end{equation}

Besides avoiding anisotropy, it was shown in~\cite{Duncan2,Endeve}
that the magnetic field is created in a very disoriented way from a
turbulent core-collapse supernovae event what  is in better
  agreement  with our present proposal than with the standard pure
  poloidal magnetic field in the $z$ direction. A limitation
of our model is that we expect that the magnetic field evolves and becomes at least partially oriented.
The oriented magnetic field is  a necessary condition to the theory of pulsars as celestial lighthouse. Here we
use the chaotic magnetic field as an approximation to better calculate the influence of the magnetic 
field in neutron star properties within the spherical symmetric TOV formalism. The stability of the chaotic
magnetic field over time is an important topic for a future work. Nevertheless it is  worth  emphasizing
that the model with a pure oriented magnetic field in the $z$ direction as normally used in the literature
\cite{Pal,Mao,Rabhi,Menezes1,Ryu,Rabhi2,Mallick,Lopes1,Dex,Benito1,Benito2,Mallick2,Ro,Dex2}
is well known to be unstable~\cite{Wright,Markey1,Markey2,Flowers}. Even though, we still gain insight and increase our knowledge
about the effects of the magnetic field on stellar matter.

\begin{figure}[tbp]
\begin{tabular}{cc}
\centering 
\includegraphics[width=.33\textwidth,origin=c,angle=270]{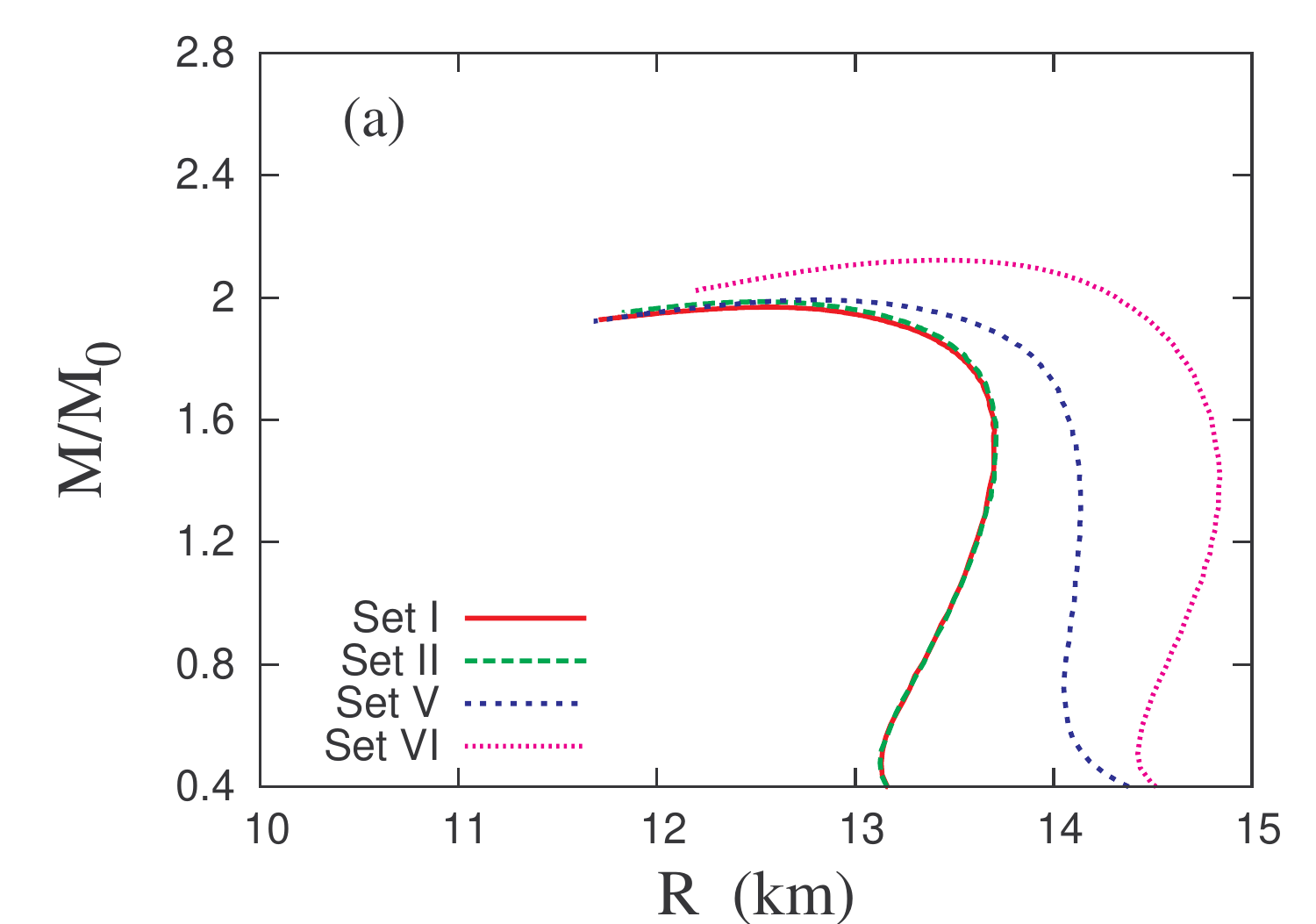} &
\includegraphics[width=.33\textwidth,origin=c,angle=270]{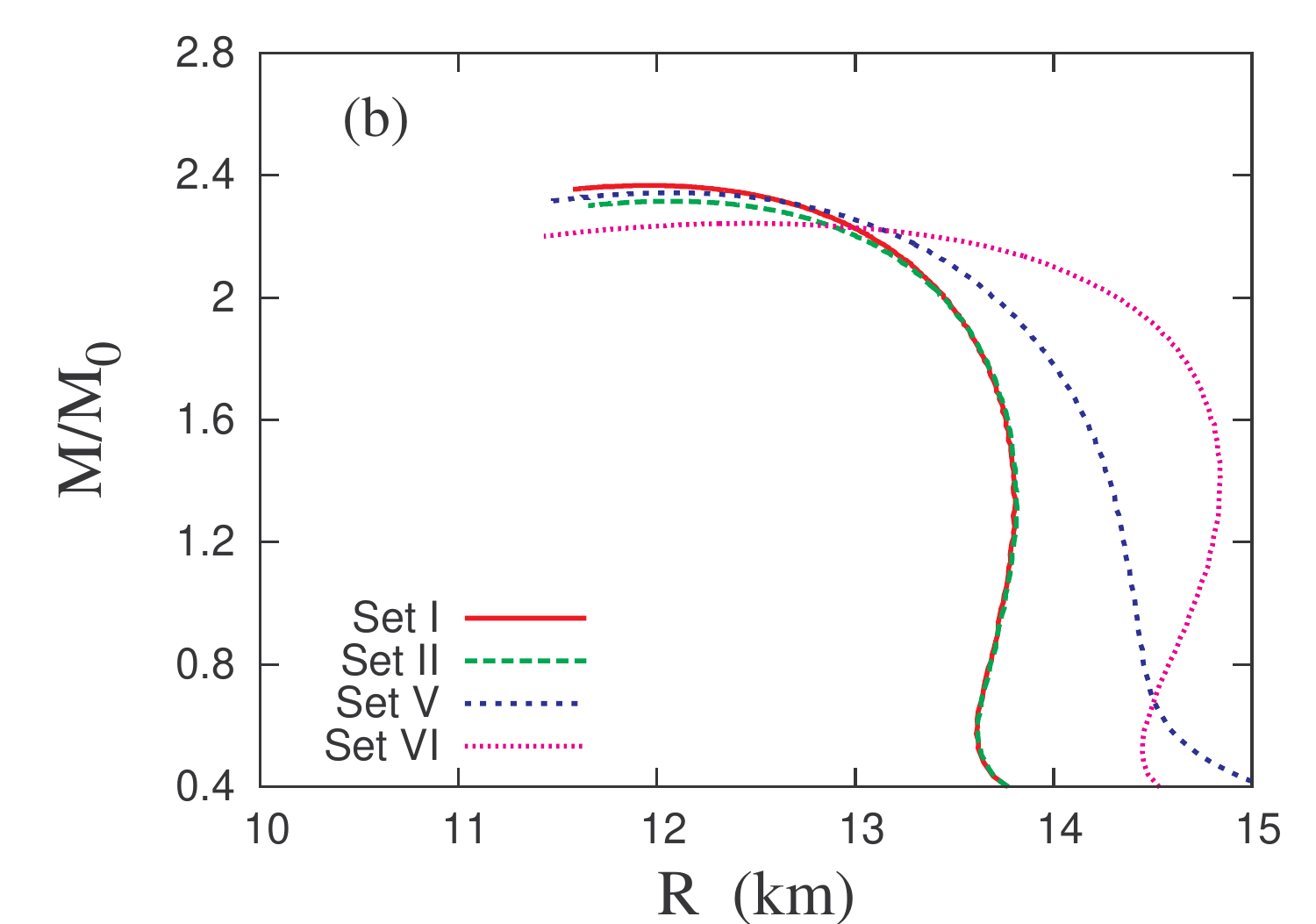}\\
\includegraphics[width=.33\textwidth,origin=c,angle=270]{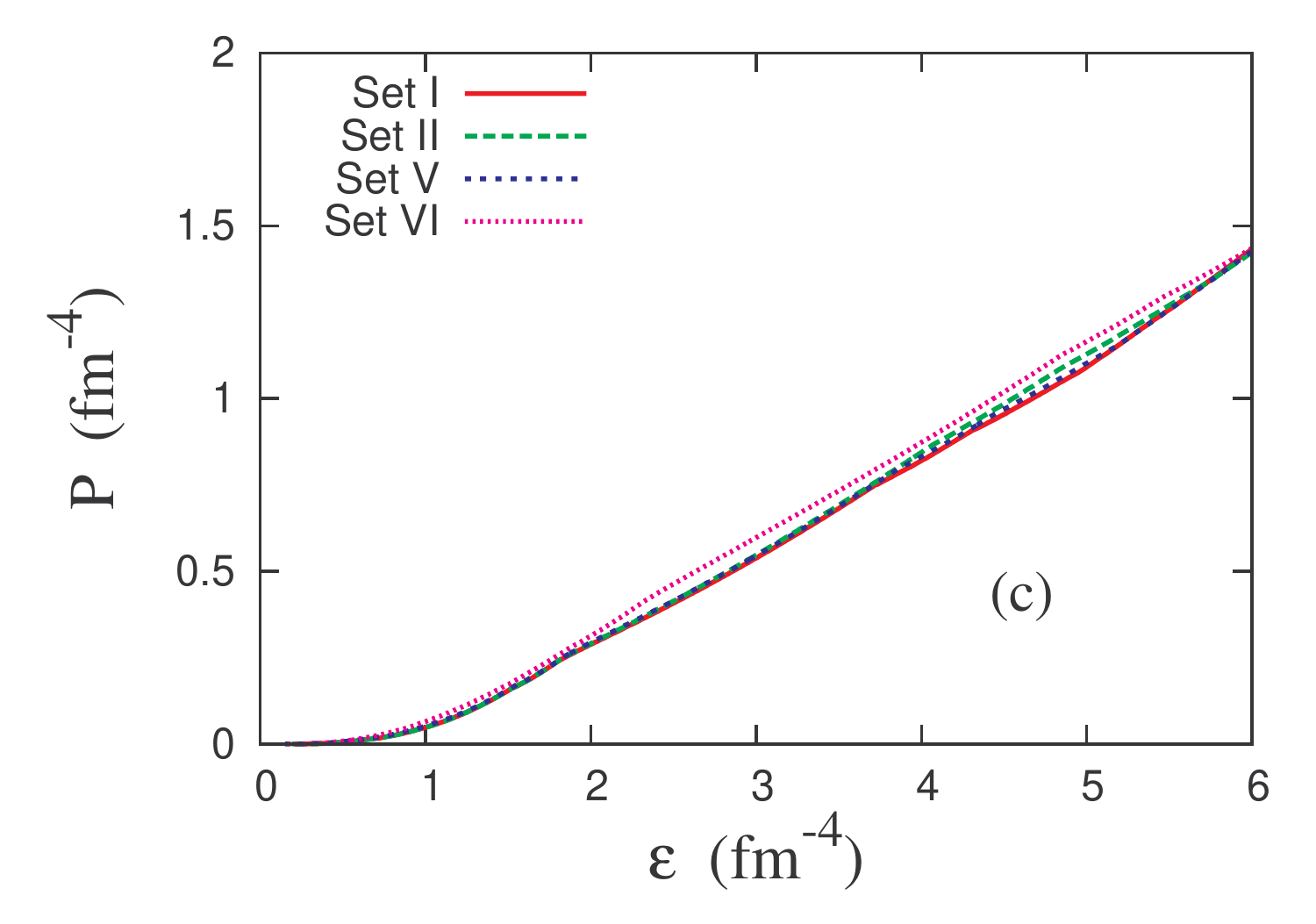} &
\includegraphics[width=.33\textwidth,origin=c,angle=270]{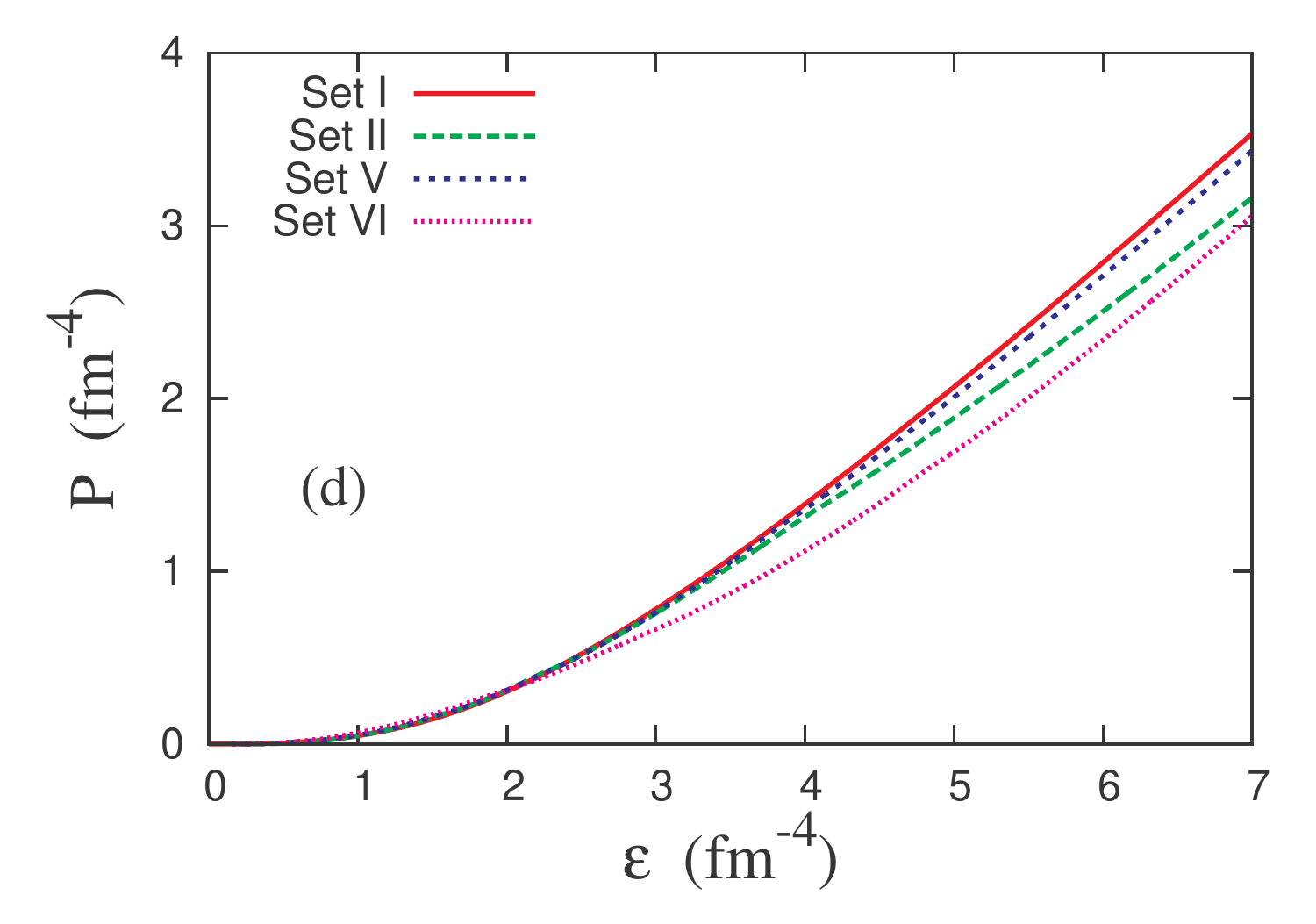}\\
\end{tabular}
\caption{\label{F2} (Color online)(Above) Mass-radius relation for neutron stars with (a) and without (b) hyperons,
and (below) their correspondent EoS (c) and (d) within a radiation pressure-like model.}
\end{figure}

Now we plot the mass-radius relation and their corresponding EoS in figure~\ref{F2}
within the formalism of Eq.~(\ref{s7}) and resume the main properties of the maximum
mass neutron star in Table~\ref{T3}.

\begin{table}[ht]
\begin{center}
\begin{tabular}{|c|ccc|ccc|}
\hline
& \multicolumn{3}{c|}{ Hyperons included} & \multicolumn{3}{c|}{Nucleons only} \\
\hline 
   Set  &  $M/M_\odot$ & R  (km)  & $\epsilon_c$ ($fm^{-4}$) & $M/M_\odot$ & R  (km)  & $\epsilon_c$ ($fm^{-4}$)  \\
 \hline
 I   &  1.94  & 12.5  &  5.01 & 2.37 & 12.1 & 5.69  \\
 
  II & 1.99 & 12.5 & 4.91 & 2.32 & 12.1 & 5.53 \\
 
  III & 2.00 & 12.6 & 4.82  & 2.30 & 12.0 & 5.60 \\
 
  IV  & 1.94 & 12.5 & 4.97 & 2.37 & 12.1 & 5.63 \\

 V & 1.99 & 12.8 & 4.78 & 2.34 & 12.1 & 5.64 \\

 VI & 2.12 & 13.4 & 4.30 & 2.25 & 12.5 & 5.37 \\ 
  
 VII & 1.99 & 12.5 & 4.94 & 2.30 & 12.1 & 5.49 \\
\hline
\end{tabular} 
\caption{Maximum mass, the correspondent radius and the central energy density
 within radiation pressure-like model for neutron stars with and without hyperons.} 
\label{T3}
\end{center}
\end{table}

When we utilize the radiation pressure formalism, a curious behaviour appears.
If on one hand the quantitative uncertainties about the maximum mass are
significantly reduced ($\Delta M < 0.18M_\odot$), qualitatively,
  different configurations now display  opposite behaviours.
While the magnetic field seems to increase the mass of neutron stars with hyperons in the
core, the largest neutron stars, which have no hyperon, have lowered their maximum masses.
This softening of the EoS { have already been} related before for strong magnetic fields~\cite{Pal,Pal0}.

\subsection{Magnetic field coupled to the energy density}

\begin{figure}[tbp]
\begin{tabular}{cc}
\centering 
\includegraphics[width=.33\textwidth,origin=c,angle=270]{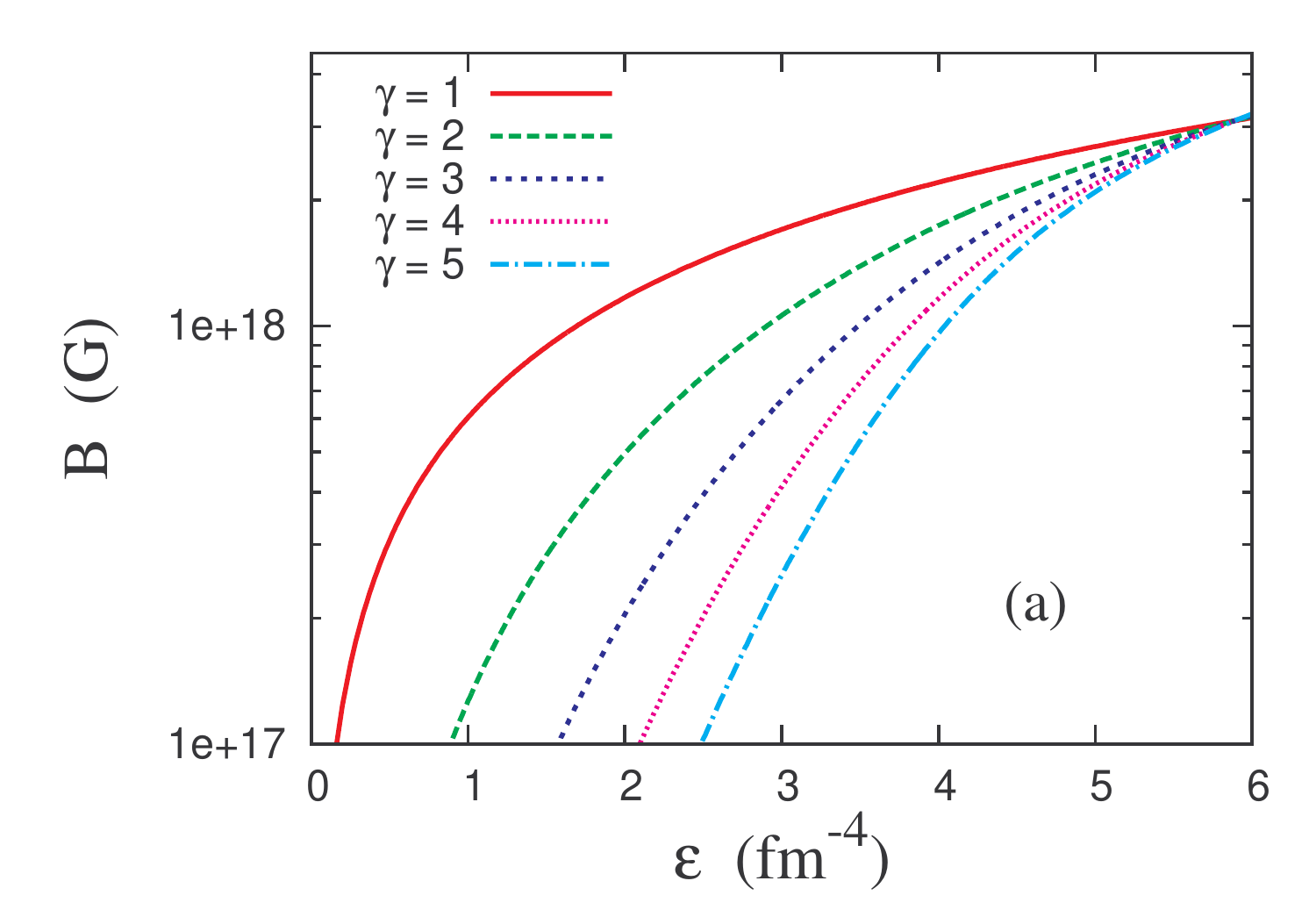} &
\includegraphics[width=.33\textwidth,origin=c,angle=270]{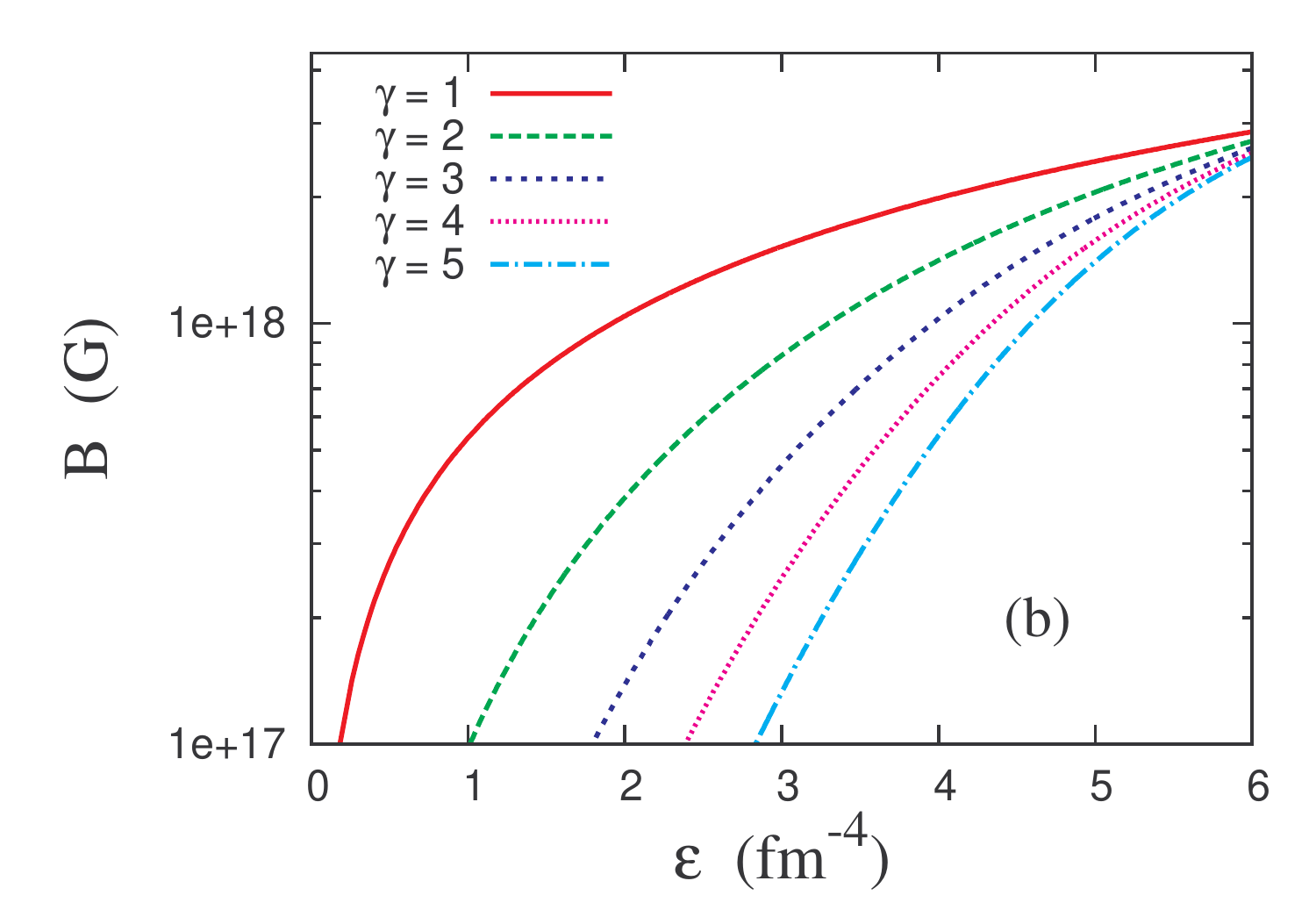}\\
\end{tabular}
\caption{\label{F3} (Color online) Magnetic field as a function of  total energy density
for neutron stars with (a) and without (b) hyperons}
\end{figure}

Another concern is about the ambiguities of the different sets of $\beta$ and $\theta$.
The first point is that, since is the energy density and not the number density that are
relevant in the TOV equations to calculate the macroscopic quantities,
 it is more natural { to use} $\epsilon$ instead of $n$.
The second point is to try to construct a model that reduces the number of free parameters.
We postulate:

\begin{equation}
B =   B_0 \bigg ({\frac{\epsilon_M}{\epsilon_c}} \bigg )^{\gamma} + B^{surf}, \label{s8}
\end{equation}
where  $\epsilon_c$  is the energy density at the centrer of the maximum mass neutron star
with zero magnetic field and $\gamma$ is any positive number, reducing the number of free 
parameters from two to only one.
$B_0$ is the fixed value of magnetic field, in our case $3.1 \times 10^{18}G$.
Also, in this way the magnetic field is no longer fixed for all neutron star configuration.
Each EoS produces a different value for $\epsilon_c$ that enters in  Eq.~(\ref{s8}).
For our particular case, $\epsilon_c = 5.01 fm^{-4}$ for neutron stars with hyperons
and $\epsilon_c = 5.69 fm^{-4}$ for neutron stars without hyperons in the core, ensuring 
 that the magnetic field does not exceed  $B_0$. 
We study the effects of the energy density-dependent magnetic field for $\gamma$ varying
from 1 to 5. We plot in figure~\ref{F3} how the magnetic field varies with the total energy density
($\epsilon_M + B^2/8\pi$) for different values of $\gamma$.

We see that for lower values of $\gamma$ the contribution of the magnetic field
is stronger. This is expected, since for $\epsilon < \epsilon_c$, the
term $\epsilon/\epsilon_c$ is lower than 1. We also see that for energy densities
around $\epsilon_c$ the value of the magnetic field is about $2.0 \times 10^{18}G$,
 more than 30\%  bellow the limit of $3.1 \times 10^{18}G$.

\begin{figure}[tbp]
\begin{tabular}{cc}
\centering 
\includegraphics[width=.33\textwidth,origin=c,angle=270]{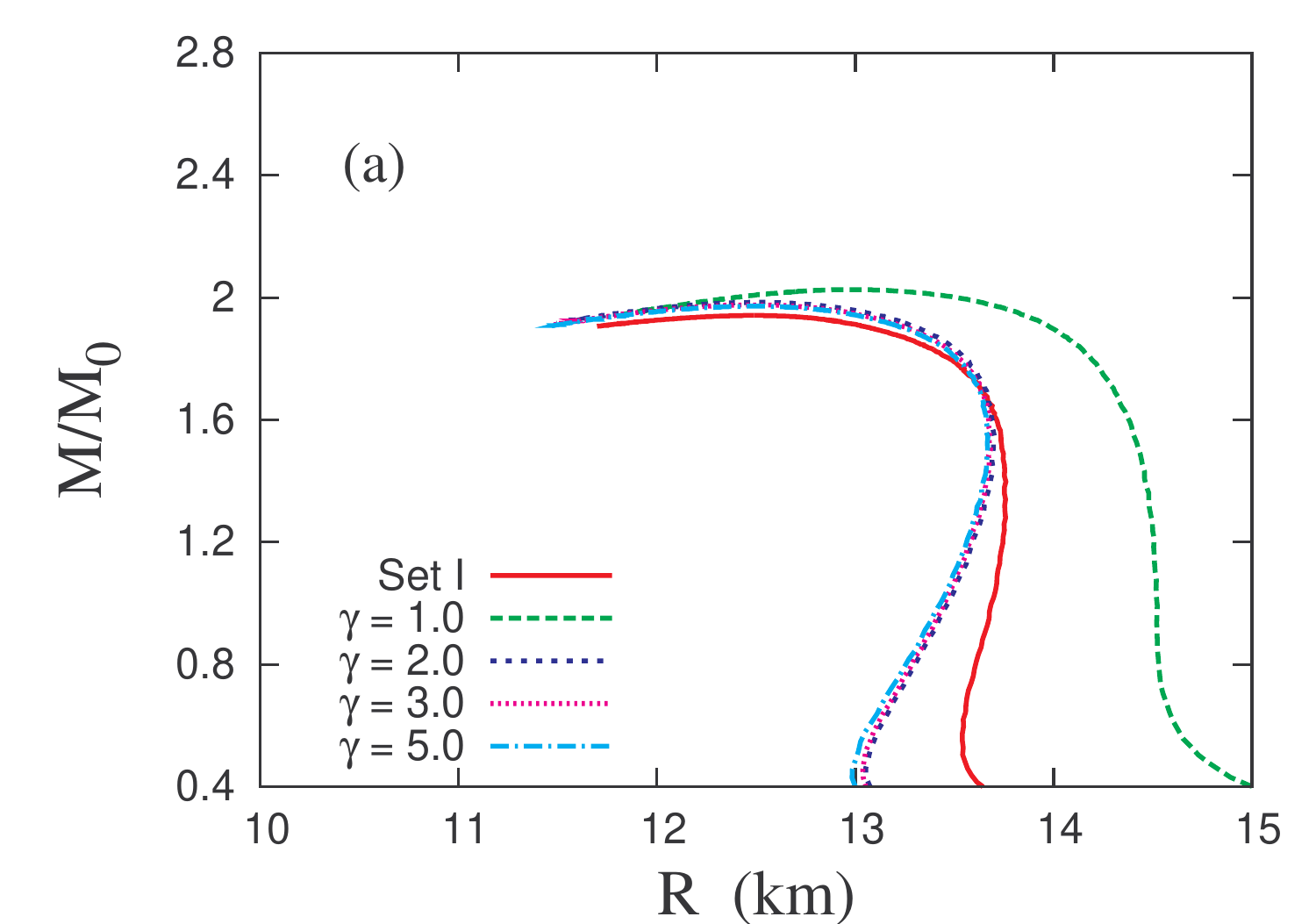} &
\includegraphics[width=.33\textwidth,origin=c,angle=270]{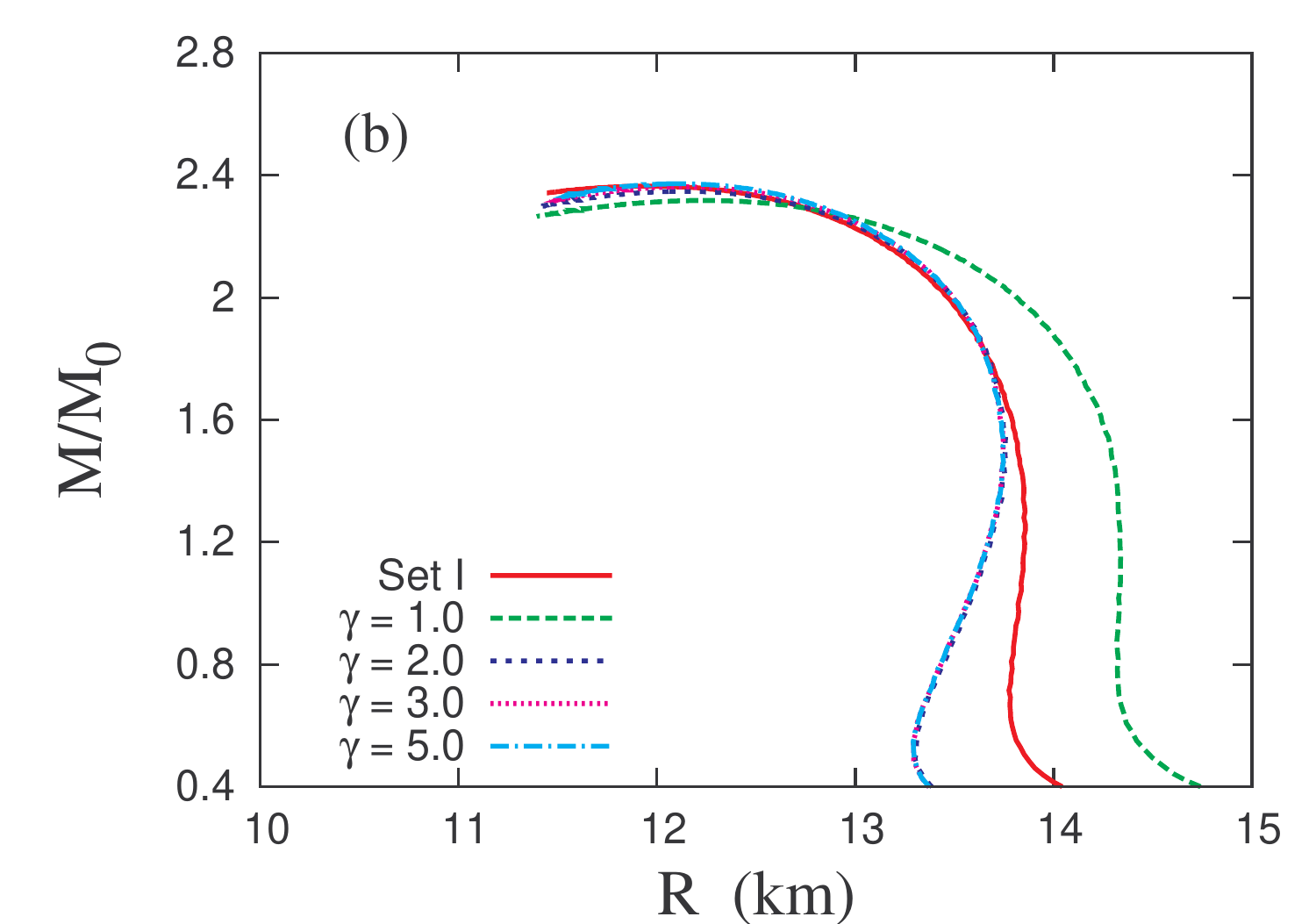}\\
\includegraphics[width=.33\textwidth,origin=c,angle=270]{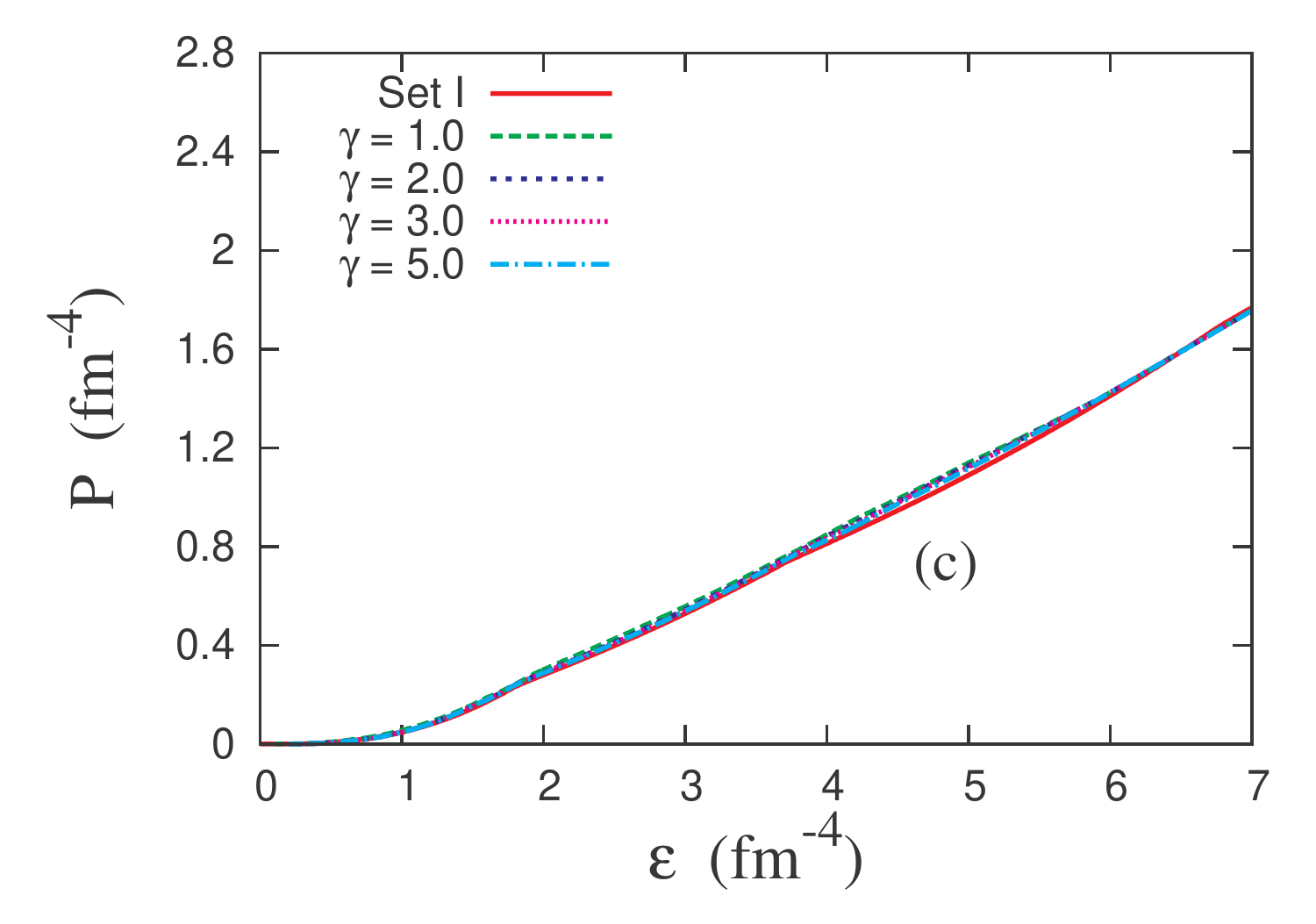} &
\includegraphics[width=.33\textwidth,origin=c,angle=270]{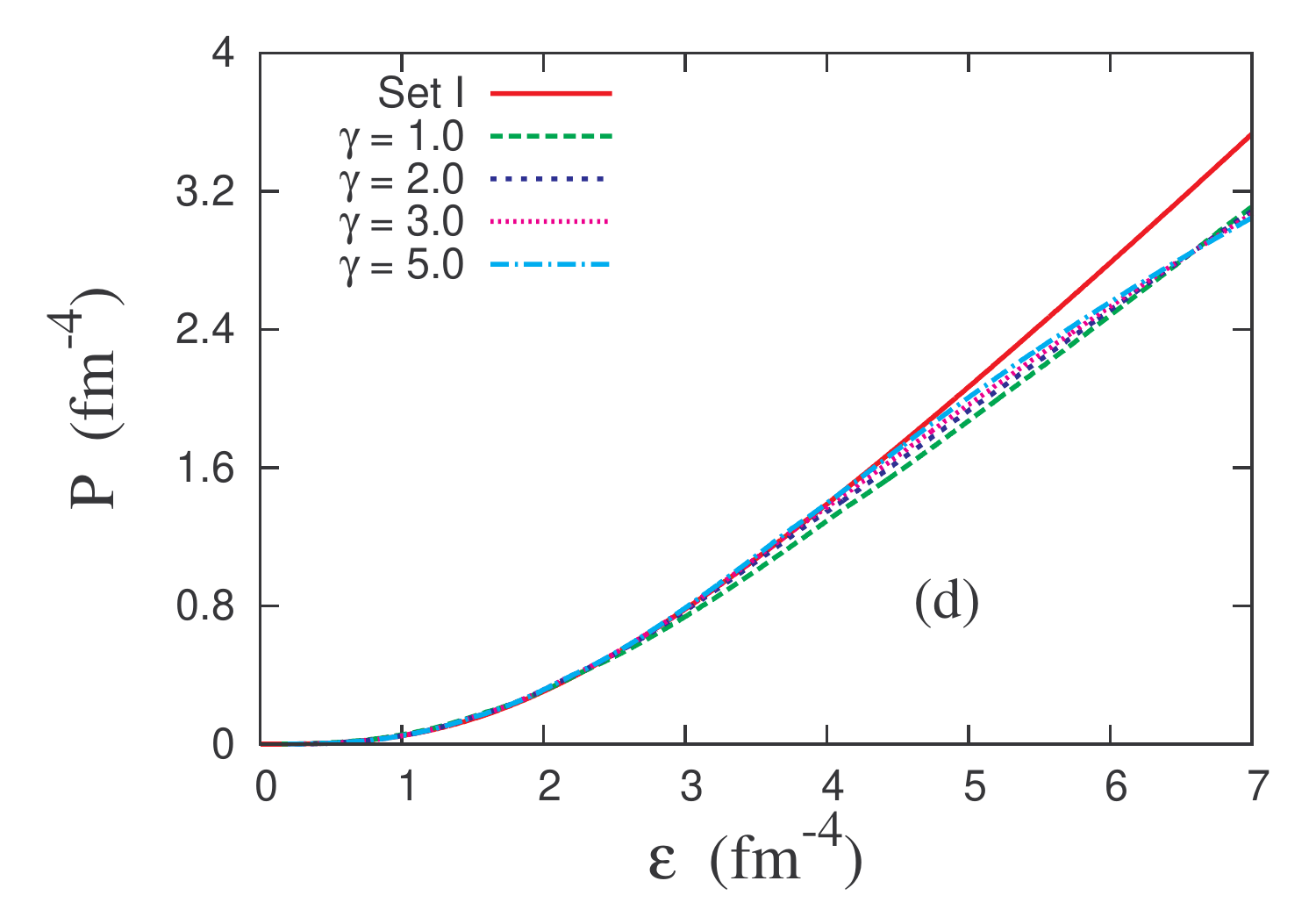}\\
\end{tabular}
\caption{\label{F4} (Color online)(Above) Mass-radius relation for neutron stars with (a) and without (b) hyperons,
and (below) their correspondent EoS (c) and (d) within energy density-dependent magnetic field.}
\end{figure}

Now we plot the influence of the energy density-depend magnetic field within the chaotic field approximation in
figure~\ref{F4} and resume the main properties of the maximum mass neutron star
in Table~\ref{T4} for different values of $\gamma$.

\begin{table}[ht]
\begin{center}
\begin{tabular}{|c|c c c |c c c|}
\hline
& \multicolumn{3}{c|}{ Hyperons included} & \multicolumn{3}{c|}{Nucleons only} \\
\hline 
   $\gamma$  &  $M/M_\odot$ & R  (km)  & $\epsilon_c$ ($fm^{-4}$) & $M/M_\odot$ & R  (km)  & $\epsilon_c$ ($fm^{-4}$)  \\
 \hline
 B = 0   &  1.94  & 12.5  &  5.01 & 2.37 & 12.1 & 5.69  \\
 
  1.0 & 2.02 & 13.0 & 4.69 & 2.32 & 12.2 & 5.49 \\
 
  2.0 & 1.98 & 12.5 & 4.92  & 2.35 & 12.1 & 5.46 \\
 
  3.0  & 1.98 & 12.5 & 4.94 & 2.36 & 12.1 & 5.44 \\

 4.0 & 1.97 & 12.5 & 4.94 & 2.37 & 12.1 & 5.46 \\

 5.0 & 1.97 & 12.5 & 4.91 & 2.37 & 12.1 & 5.49 \\ 
\hline  
\end{tabular} 
\caption{Maximum mass, the correspondent radius and the central energy density
 within energy density-dependent magnetic field for neutron stars with and without hyperons.} 
\label{T4}
\end{center}
\end{table}

We see that with the energy density-dependent magnetic field our { uncertainty} drops dramatically. 
When we introduce the chaotic magnetic field { approach}, the uncertain of the maximum mass drops from  1.08$M_\odot$
to $0.18M_\odot$. Now with energy density-dependent magnetic field, our uncertainty  reaches a maximum of
$0.05M_\odot$. Also, for $\gamma$=1.0, the radius of the neutron stars grows, while for $\gamma$ $\ge$ 2.0
the radius decreases. For instance, for the canonical 1.4$M_\odot$, in the absence of the magnetic field. the radius
is 13.8 km. Utilizing the energy density-dependent magnetic field, for $\gamma$ = 1.0, the radius becomes 14.3 km,
however, for $\gamma$ $\ge$ 2.0 the radii of the canonical neutron
star masses are all equal to 13.6 km, a value 
lower than
in the zero magnetic  field approximation. Nevertheless, since 
there is no significant variation in the values of the maximum masses,  
in this regime ($\gamma$ $\ge$ 2.0) we are able to construct a true free parameter model!
Notice, however that { in} case  we { had} used Eq.(~\ref{s4})
rather than Eq.~(\ref{s7}), even with the new  prescription
for the magnetic field, the maximum stellar mass { would} once
again be parameter dependent, 
varying from 2.15$M_\odot$
for $\gamma$ = 2.0 to 2.03$M_\odot$ to $\gamma$ = 5.0, enhancing the importance of the use of the chaotic 
magnetic field approximation.

Although the formalism derived from Eq.~(\ref{s7}) seems more suitable than the one obtained from Eq.~(\ref{s4}),
the matter of how the magnetic  field varies in the neutron star
interior is still unknown. 
{ Notice that } sets II to VII in Table~\ref{T3} and all values of $\gamma$ 
in Table~\ref{T4} are, at first, all on equal footing. Therefore, to constrain the variation of the magnetic field in the neutron star interior
additional information is required. Recent progress on neutron star radii are seen
in both, theoretical and observational
methods~\cite{Hebeler,Lim,Steiner2}. Hence, our proposal { given} in Eq.~(\ref{s8}) for $\gamma$ $\ge$ 2.0
seems more suitable, since it predicts lower radii, in both configuration, with and without hyperons, reducing
the radii even when compared with those obtained at zero magnetic field. Moreover, based on a chiral
effective theory, ref.~\cite{Hebeler} constrains the radius of the canonical 1.4 $M_\odot$ neutron star to 9.7 - 13.9 km.
In this case, as seen from figure~\ref{F2}, sets V and VI have to be ruled out, so as $\gamma$ = 1.0, once the canonical neutron star show 
 a radius beyond the theoretical limit. 

 In our work we utilize the TOV equations to obtain the mass-radius relation. This is allowed
 due to the use of the chaotic magnetic field approximation,  which
 avoids the anisotropy features.  It is worth mentioning that there are
in the literature more sophisticated calculations that consider the anisotropy in solving Einstein's field equations
in axisymmetric regime, either via fully general relativistic
formalism~\cite{Bocquet,Cardall,Micaela} or  by expanding the metric
around the spherically symmetric case~\cite{Mallick2}. The small increase of the maximum mass that we found
fully agrees with both refs.~\cite{Micaela,Mallick2}. The great
advantage of our model is the much simpler and 
didactic way to study the effects of magnetic field, besides the
tremendous less computational price. It is also important to emphasize
that the results computed from the LORENE modified code in
\cite{Micaela}, generate a magnetic field in the central regions of the
star of the order of  $10^{18}$ G, which seems reasonable but at the
surface, its value is of the order of $10^{17}$ G, which is
too large in
contrast with observational results, which point to a maximum 
value of $10^{15}$ G.

\section{Quark Star Formalism and Higher Magnetic Fields}

Now we discuss  compact  stars which are
composed not by hadrons but by free quarks. This proposal
relies on the  Bodmer-Witten conjecture~\cite{Bodmer,Witten}, that strange matter
may be the actual ground state of baryon matter at high densities.
Unlike the hadronic neutron stars, quark stars are bound by the strong force,
instead of the gravitational one. Due to this fact, in quark stars the strength of the magnetic
field can reach values far above those found in conventional neutron stars.
 According to ref.~\cite{Ferrer}, the magnetic field in quark stars can reach
values up to $10^{20}$G.

This fact reinforces the importance of the chaotic magnetic field formalism.
For magnetic fields above $3.1\times 10^{18}G$ the anisotropy is 
  no longer negligible as pointed out in ref.~\cite{Xu1,Xu2}. So, the chaotic
magnetic field is the only reasonable theory  that allows the
study of magnetized compact stars for high magnetic fields in the TOV context,
once its predicts a truly isotropic EoS.  Next, we also keep the standard formalism
based in eq.~(\ref{s4}) for comparison.

To study quark matter we use the MIT bag model~\cite{Chodos}.
Magnetized quark stars within this context have already been exhaustively discussed
in the literature~\cite{Xu1,Dex2,Ferrer,Gonzales,Laura,Manresa}. 

The Lagrangian~\cite{KJ} of the MIT bag model in the presence of 
an external magnetic field reads:

\begin{equation}
\mathcal{L}_{MIT} = \bigg \{ \sum_q  \bar{\psi}_q \bigg [ \gamma^{\mu} (i \partial_\mu - e_qA_\mu) - m_q  \bigg ]\psi_q  - B \bigg \}\Theta(\bar{\psi}\psi)
\end{equation}
where the sum runs over the three lightest quarks: up, down and strange; $\psi_q$ are the quark Dirac fields and $m_q$ and $e_q$ are the 
quarks masses and charges respectively.
The model is very simple, and consider that quarks are free fermions inside a bag. All the information about the strong
force { comes from the bag pressure or vacuum pressure, $B$; and
 $\Theta(\bar{\psi}_q\psi)$ is the Heaviside step function to assure that the quarks exist only confined to the bag.
 
 The thermodynamic relations are the same as for the leptons plus the bag constant $B$, that yields a plus sign in the 
 energy density and a minus sign in the pressure. Leptons are added in order to achieve chemical equilibrium.
The parameters of the model needs to satisfies the necessary conditions that ensure that quark matter is the true ground
state: two-flavor quark matter must be unstable (i.e., at zero temperature its energy per baryon has to be larger
than 930 MeV, the iron binding energy) and the three-flavor quark matter must be stable (i.e., its energy per
baryon must be lower than 930 MeV, also at T = 0). Values inside the so called stability windows fulfill these constraints~\cite{James},
and to accomplish that, here we use $m_q = m_u = 5~MeV, m_s = 135~MeV$ and $B = 150~MeV^{1/4}$.

We compare the effects of utilizing chaotic magnetic field instead of the
standard eq.~(\ref{s4}); and in both cases we utilize the energy density-dependent magnetic field of eq.~(\ref{s8}),
and $\gamma = 4$. We choose three values of magnetic field: $3\times 10^{18}G$, $6\times 10^{18}G$
 and $9\times 10^{18}G$, although fields up to $10^{20}G$ are possible, but not probable.
We plot the mass-radius relation and their corresponding EoS in figure~\ref{Fq}
and resume the main properties of the maximum quark star mass  in Table~\ref{Tq}.
With the standard formalism, the quark star maximum mass increase
up to $17\%$, while with the chaotic magnetic field approximation the maximum increase
is around $8\%$. Again, this low increases of the mass agrees with more sophisticated calculations~\cite{Bocquet,Cardall,Micaela,Mallick2}.

\begin{table}[ht]
\begin{center}
\begin{tabular}{|c|c c c |c c c|}
\hline
& \multicolumn{3}{c|}{ $P_T = P_M + B^2/8\pi$} & \multicolumn{3}{c|}{$P_T = P_M + B^2/24\pi$} \\
\hline 
   $B_0 (\times 10^{18}G)$  &  $M/M_\odot$ & R  (km)  & $\epsilon_c$ ($fm^{-4}$) & $M/M_\odot$ & R  (km)  & $\epsilon_c$ ($fm^{-4}$)  \\
 \hline
 B = 0   &  1.73  & 9.6  &  7.25 & 1.73 & 9.6 & 7.25  \\
 
  3.0 & 1.80 & 9.4 & 7.90 & 1.76 & 9.7 & 7.14 \\
 
  6.0 & 1.91 & 9.6 & 7.80  & 1.80 & 10.0 & 6.78 \\
 
  9.0  & 2.03 & 10.0 & 7.34 & 1.88 & 10.3 & 6.30 \\
\hline  
\end{tabular} 
\caption{Maximum mass, the correspondent radius and the central energy density
 for different values of magnetic field.} 
\label{Tq}
\end{center}
\end{table}
\begin{figure}[tbp]
\begin{tabular}{cc}
\centering 
\includegraphics[width=.33\textwidth,origin=c,angle=270]{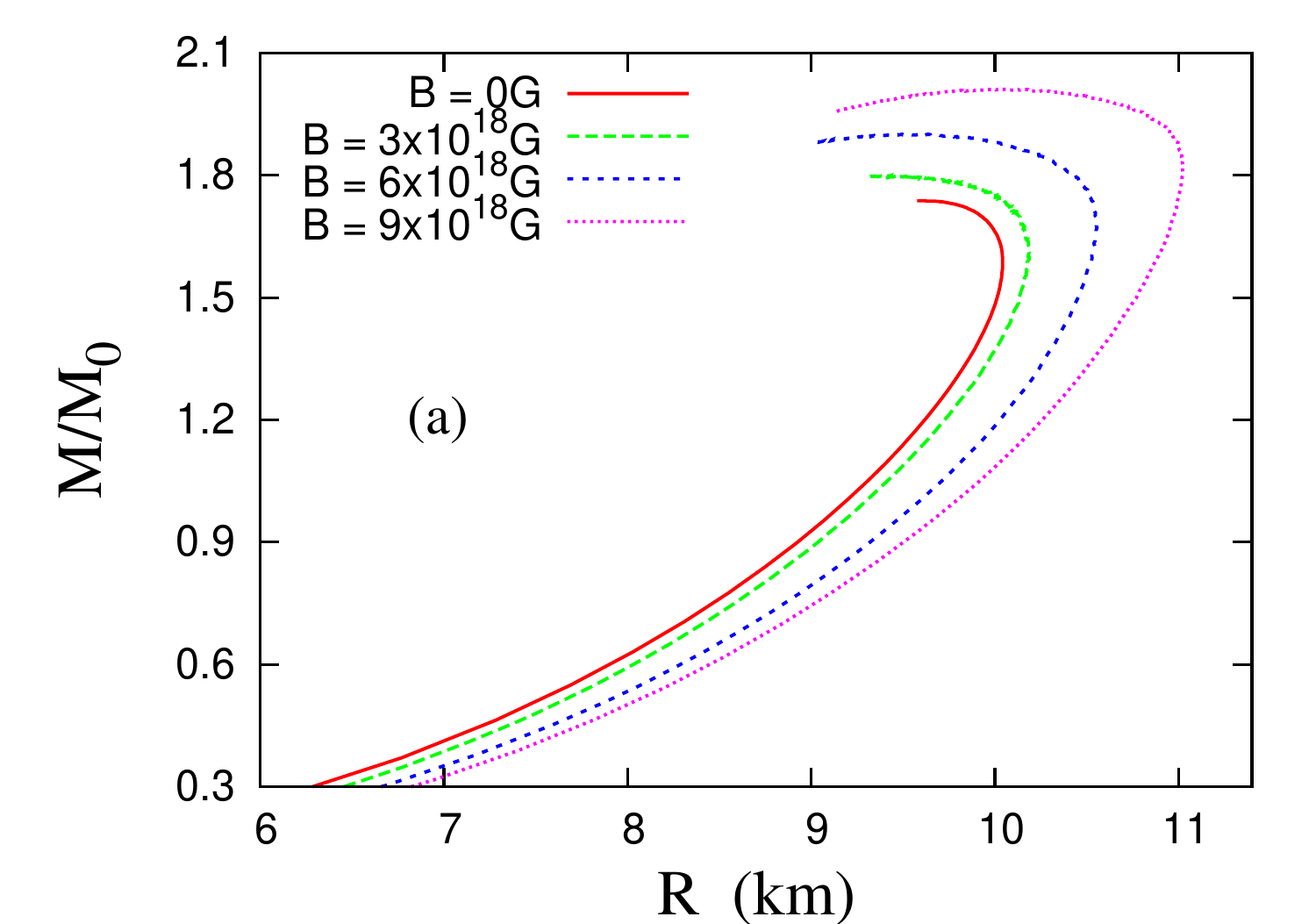} &
\includegraphics[width=.33\textwidth,origin=c,angle=270]{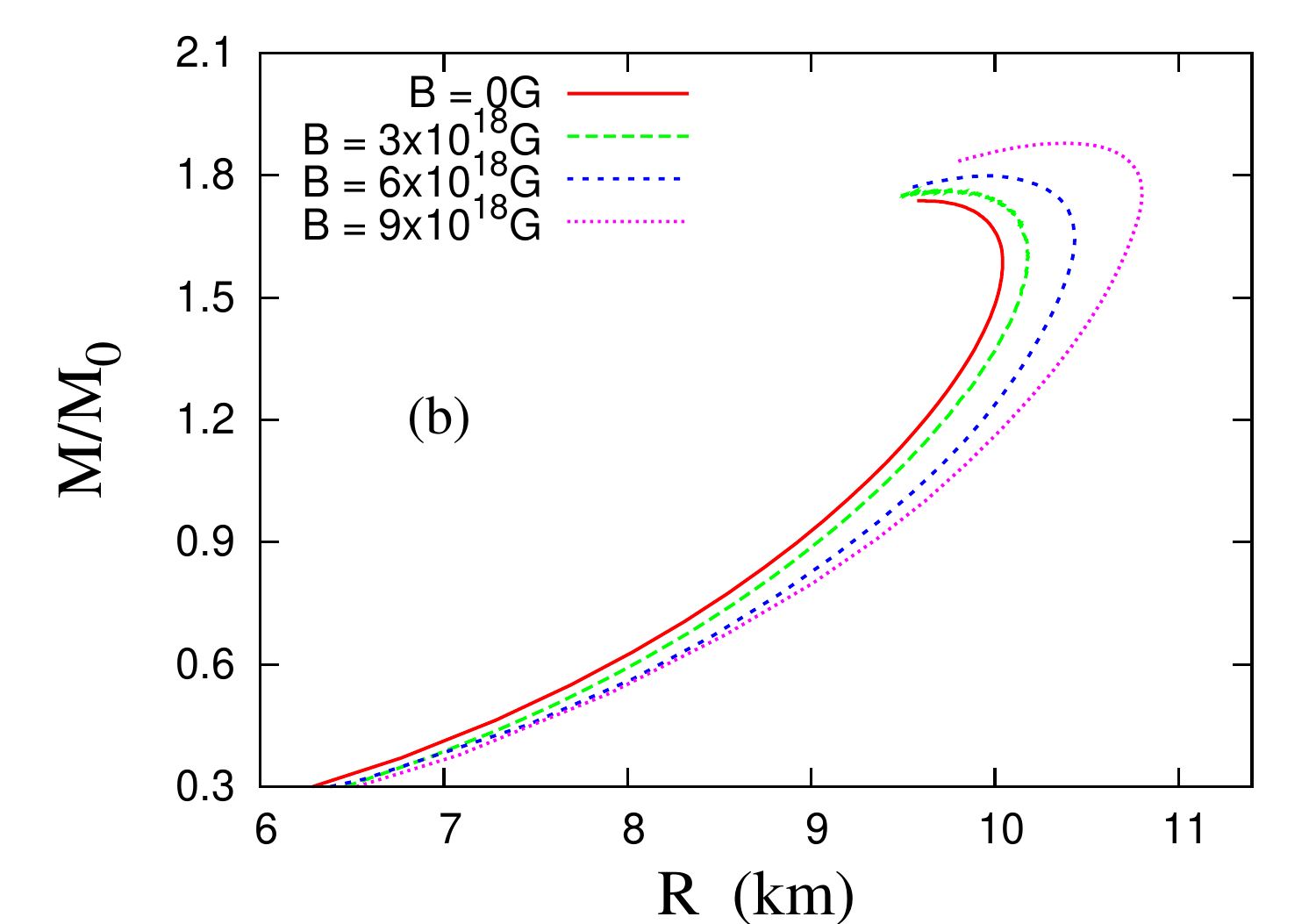}\\
\includegraphics[width=.33\textwidth,origin=c,angle=270]{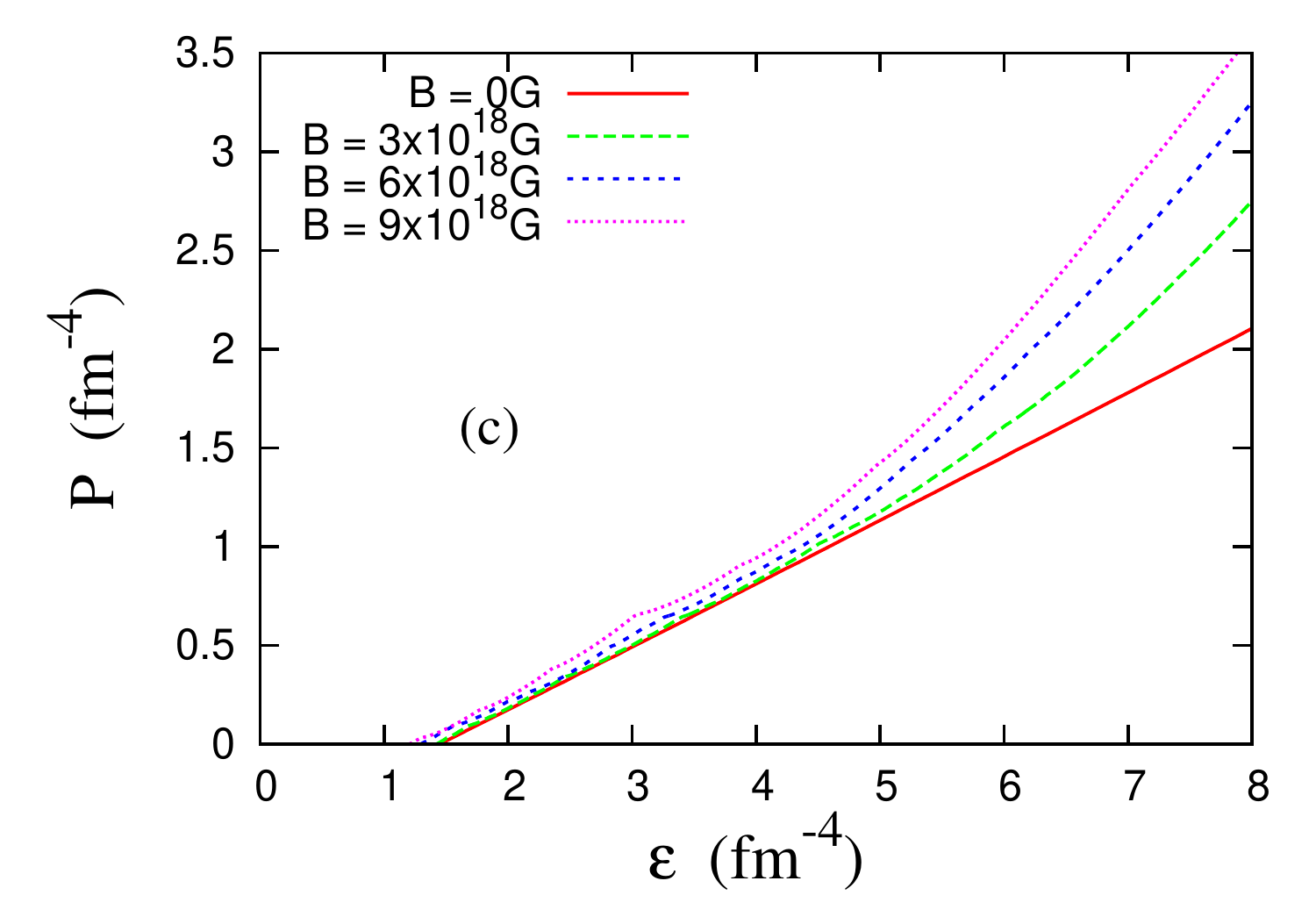} &
\includegraphics[width=.33\textwidth,origin=c,angle=270]{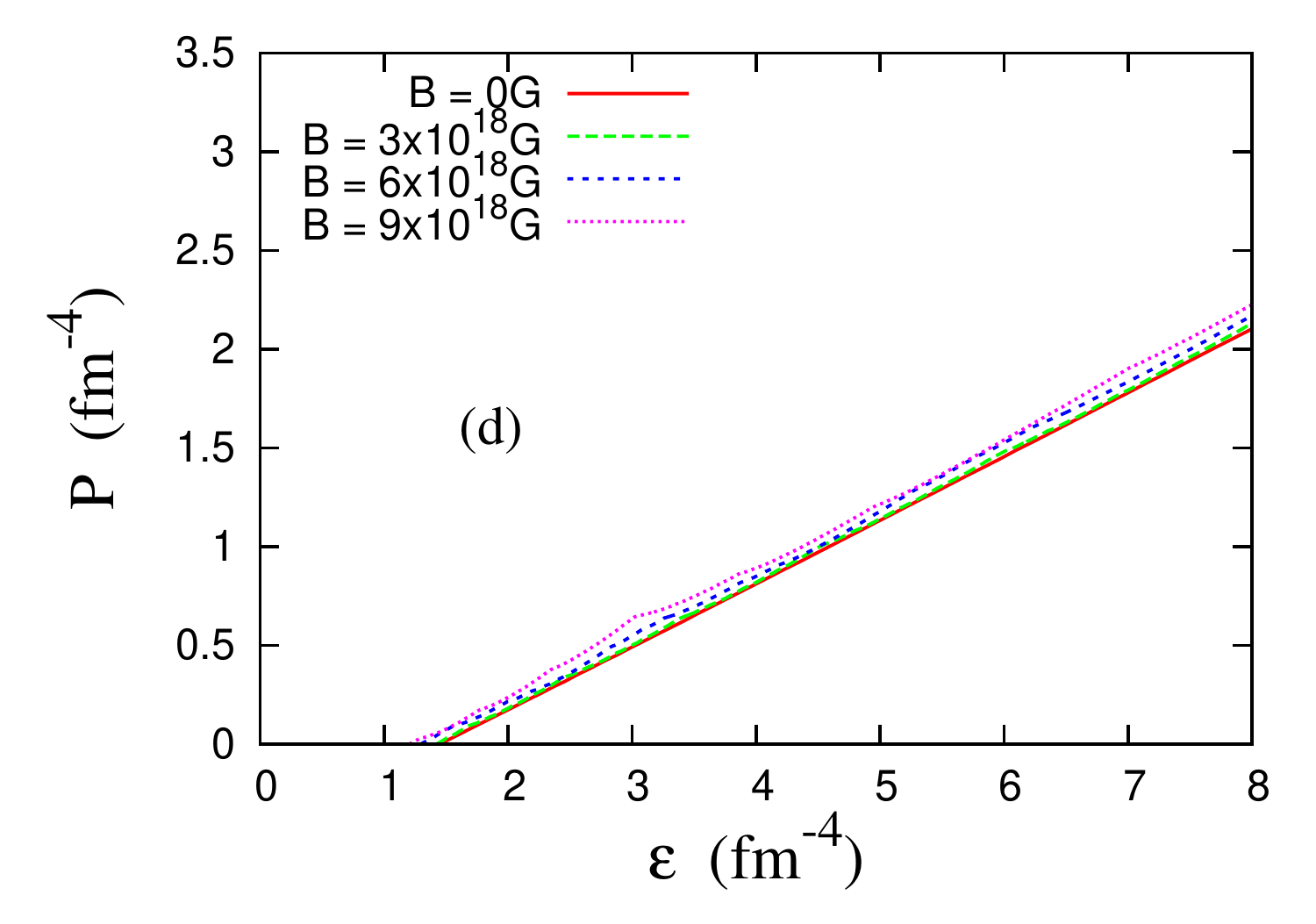}\\
\end{tabular}
\caption{\label{Fq} (Color online)(Above) Mass-radius relation for quark stars within (a) standard $P_B$ = $B^2/8\pi$ and (b)
within chaotic magnetic field approximation, and (below) their correspondent EoS (c) and (d) within energy density-dependent magnetic field.}
\end{figure}

\begin{figure}
\centering
\includegraphics[width=0.45\textwidth, angle =270 ]{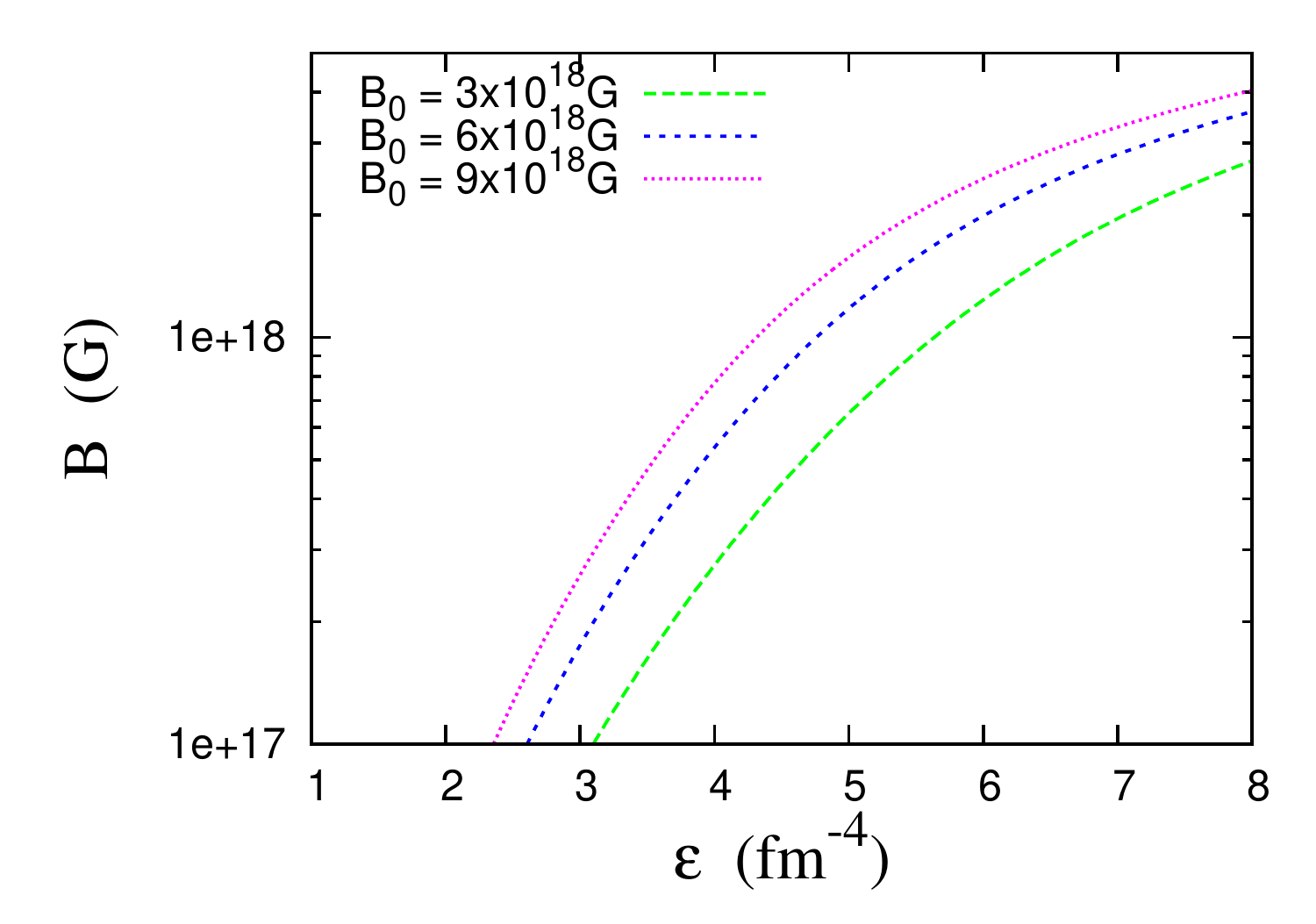}
\caption{Magnetic field as a function of  total energy density
for quark stars}
\label{figB}     
\end{figure}

These differences are more evident when we look at the EoS. In the standard approach, the pressure
significantly increases with the magnetic field, while in the chaotic
magnetic field, all of the EoS present
just a low hardening with the increase of the magnetic field.
 We see that unlike neutron stars, the radii of quark stars increase with the magnetic field.
 This is due to the fact that neutron stars have a crust, while quark stars do not.

 As far as the two solar masses neutron stars are concerned~\cite{Demorest,Antoniadis},
previously ruled out MIT EoS could be restored in the context of the  standard approach for the magnetic field, 
once the maximum mass would reach 2.03$M_\odot$. However this is not possible if the chaotic magnetic field is used,
 since in this case the increase of the mass is  relatively small, even for very high magnetic fields.
 One can also ask about the validity of the results, since from Table~\ref{Tq} the central 
energy density is higher for magnetized quark stars than for the non-magnetized ones. However, the central
energy density is the sum of the matter energy density plus the contribution of the magnetic field
($\epsilon_T = \epsilon_M + B^2/8\pi$). From figure~\ref{figB} we see that the magnetic field at central
energy density for $B_0 = 3\times 10^{18}G$, $6\times 10^{18}G$ and $9\times 10^{18}G$ are
 $2.6\times 10^{18}G$, $3.6\times 10^{18}G$ and $4.0\times 10^{18}G$
 respectively and therefore, never exceed the value of $B_0$.

To finish our analyses we discuss the limitations of our model. Although the chaotic magnetic field gives us
a { more suitable} interpretation of the magnetic field pressure,
and the energy density-dependent magnetic
 field give us a model free parameter,
there is still the fact that a variable magnetic field violates one of Maxwell's equations, since the divergent
of the magnetic filed is { no longer zero}. This is a problem present in all models with variable magnetic
field~\cite{Pal,Mao,Rabhi,Menezes1,Ryu,Rabhi2,Mallick,Lopes1,Dex,Benito1,Benito2,Mallick2,Ro,Dex2}. Nevertheless
it is worth { bering} in mind that we are dealing with approximations and a variable magnetic field agrees with the scalar Viral theorem~\cite{Teu}.

\section{Conclusion}

In this work we review the current formalism used to consider  magnetic fields on neutron star properties. 
We see that  the ambiguities found come largely from an inadequate formalism to introduce the contribution
of the magnetic field to the pressure, since it does not take into account the anisotropy on the stress tensor
~\cite{Rabhi,Menezes1,Ryu,Lopes1,Dex,Benito1,Benito2,Ro,Prakash}. To solve the problem, we choose the chaotic magnetic field, that
 skirts the matter of the anisotropy and agrees with both thermodynamic~\cite{Misner,Zel} and field theory~\cite{Serot}
concept of pressure. The first consequence of this choice is the strong decrease of the quantitative uncertainty on the influence 
of the magnetic field. While in the standard models, the magnetic field could increase the maximum mass to values
beyond 1.08$M_\odot$, in the chaotic magnetic field approximation the variation of the masses are never superior than 0.18$M_\odot$.
Constraining the radius of the neutron stars to the recent studies~\cite{Hebeler,Lim,Steiner2}, the uncertainty drops
to a maximum of $0.07M_\odot$.

Our second task is to eliminate the ambiguities of the variation of the magnetic field
 by introducing the energy density-dependent magnetic field with only one free parameter $\gamma$, in opposition
to the density dependent magnetic field that has two free parameters. Moreover, since the only acceptable values are $\gamma$ $\ge$ 2.0,
due to the radius constraint, we have prescribed a true parameter free model. Another point that is worth bearing in mind, is that since we do not know how
the magnetic field varies in the neutron star interior all
parametrizations of the magnetic field are equally valid. The
advantages of  using Eq.~(\ref{s8})
are: first, it generates a parameter free model; second, the magnetic field couples to the energy density, which is a relevant parameter 
in solving the TOV equations, in opposition to Eq.~(\ref{s4}), 
  where it couples to  the number density; third, our approach produces neutron stars with small radii, in a better agreement with the recent studies on this subject~\cite{Hebeler,Lim,Steiner2}
  and forth, our results agrees with more sophisticated
   calculations~\cite{Micaela,Mallick2}, 
however utilizing a much simpler and didactic way to study the effects
of the magnetic field and with 
much less computational price. Furthermore, it is well known that
poloidal magnetic fields are unstable.

 We also study quark stars, and in this case the importance of chaotic magnetic field is even more
evident, since it permits the study of magnetic fields that are highly
anisotropic in the standard models and for which the validity
of the TOV equations are very questionable.
We see that the increase of the maximum mass of quark stars is more
subtle in the chaotic magnetic field approach, and unlike the standard models,
ruled out EoS cannot be restored by just increasing the strength of the
magnetic field. We finish our work stating that the magnetic field up to $3.1\times 10^{18}G$ has just a subtle
contribution to the maximum mass, around $2-3\%$.
Increases of the order of $10\%$ or even higher as found in  previous works
~\cite{Menezes1,Lopes1,Dex,Benito1,Benito2,Ro},  among many others, seems rather artificial
due to a  possible inadequate choice of the formalism
to treat anisotropies. Nevertheless, besides the energy symmetry slope,  the magnetic field  plays a subtle but not
negligible role in the neutron star radii puzzle. 

\acknowledgments

 This work is partially supported by CNPq, CAPES, FAPESC and CEFET/MG.

\appendix

\section{Anisotropy and shear stress in Maxwell's stress tensor}

For a magnetic field in the $z$ direction,
the stress tensor reads:

\begin{eqnarray}
T=\left[\begin{array}{rrr}
B^2/8\pi&0&0\\
0&B^2/8\pi&0\\
0&0&-B^2/8\pi
\end{array}\right] .\label{a1}
\end{eqnarray} 

At first glance it seems that there is no shear stress, but  there are two different pressures. 
Often these different elements are called parallel and perpendicular pressures ref.~\cite{Ro,Dex2,Mallick2}:

\begin{equation}
P_{\perp} = P_M + \frac{B^2}{8\pi}; \quad P_{||}  = P_M -  \frac{B^2}{8\pi} ,\label{a2}
\end{equation}
where $M$ stands for the matter.

However as noted in ref.~\cite{Micaela} these ``pressures'' do not correspond to the thermodynamical pressure.
A simple way of understanding this fact is remembering that the pressure is a scalar and cannot depend on the
direction!

Let us explore deeper the problem. Instead of  choosing the magnetic field along the $z$ axis, we can choose
a magnetic field $\vec{B} = \alpha B_0\hat{x} + \beta B_0\hat{y} + \gamma B_0\hat{z}$, indicating an arbitrary strength in the $x$, $y$
and $z$ directions. In this case, the total energy density is:

\begin{equation}
\epsilon = \frac{B^2}{8\pi} = \frac{B_0^2}{8\pi} \cdot (\alpha^2 + \beta^2 + \gamma^2), \label{a3}
\end{equation}
the stress tensor reads: 

\begin{eqnarray}
T'=\frac{B_0^2}{8\pi}\left[\begin{array}{rrr}
(\beta^2 + \gamma^2 - \alpha^2)&-2\alpha\beta&-2\alpha\gamma\\
-2\alpha\beta&(\alpha^2 +\gamma^2 -\beta^2)& -2\beta\gamma\\
-2\alpha\gamma&-2\beta\gamma&(\alpha^2 + \beta^2 - \gamma^2)
\end{array}\right] \nonumber
\end{eqnarray}

This could seem a different problem but it is not. This is just a passive rotation
in the coordinate system where the new stress tensor $T'$ is a rotation of
the old stress tensor $T$.

\begin{equation}
T' = RTR^T, \label{a4}
\end{equation}

For a passive rotation we have~\cite{Sakurai} :

\begin{eqnarray}
R =\left[\begin{array}{rrr}
\cos\phi & \sin\phi\cos\theta&\sin\phi\sin\theta\\
-\sin\phi&\cos\phi\cos\theta&\cos\phi\sin\theta\\
0& -\sin\theta & \cos\theta
\end{array}\right] \label{a5}
\end{eqnarray}

Now we can write $\alpha$, $\beta$ and $\gamma$ in terms 
of $\theta$ and $\phi$:

$$\left\{\begin{array}{r}
\alpha \quad \mbox{=}\quad \sin\theta\sin\phi\\
\beta \quad \mbox{=}\quad \sin\theta\cos\phi\\
\gamma \quad \mbox{=} \quad \cos\theta\\
\alpha^2 +\beta^2 +\gamma^2 \quad = \quad 1
\end{array}\right.
$$

Let us study  two examples. First, 
we make a passive rotation in such a way that:
$\alpha$ = $1/\sqrt{3}$, $\beta$ =0 and $\gamma$ = $\sqrt{2/3}$.
In this case:

\begin{eqnarray}
T'=\frac{B_0^2}{8\pi}\left[\begin{array}{rrr}
1/3 & 0 & -2\sqrt{2}/3\\
0 & 1 & 0 \\
-2\sqrt{2}/3 & 0 & -1/3
\end{array}\right]. \label{a6}
\end{eqnarray} 

Now, instead of two, it seems that there are three different {\it pressures}.
Also, a simple rotation reveals  the presence of shear stress,
that was hidden due to a particular choice of the coordinate system
in eq. (\ref{a1}).

A second choice is to perform a passive rotation in such way that:
$\alpha$ = $1/\sqrt{3}$, $\beta$ =$1/\sqrt{3}$ and $\gamma$ = $1/\sqrt{3}$,
and then

\begin{eqnarray}
T'=\frac{B_0^2}{8\pi}\left[\begin{array}{rrr}
1/3 & -2/3 & -2/3\\
-2/3 & 1/3 & -2/3 \\
-2/3 & -2/3 & 1/3
\end{array}\right] \label{a7}.
\end{eqnarray}

It seems that we have solved the problem of anisotropy, but
we gained additional  shear stress terms.

These examples could lead us to the absurd conclusion that the pressure could
depend not only on the direction but also on the coordinate system.

Since the pressure is a scalar, we need an invariant way to calculate it.
The natural choice is calculate the magnetic pressure in the same way we calculate
the pressure of the strong interacting matter. As pointed in ref.~\cite{Serot}:

\begin{equation}
P = \frac{1}{3}< T_i^i>, \nonumber
\end{equation}
that yields P = $\epsilon$/3, no matter if we choose eq. (A.1), (A.6) or (A.7)  for the stress
tensor.

However, although we can correctly predict the value of  P = $\epsilon$/3, to obtain the 
thermodynamical pressure~\cite{Misner,Zel} we need that all components of
the stress tensor be equal.

So we introduce the chaotic magnetic field. This allow us to correctly predict the pressure P = $\epsilon$/3, which is 
truly independent of the direction and the coordinate system,
 restoring the concept of thermodynamical  pressure,  and avoiding all the problem related to the  shear stress
and anisotropies.


\end{document}